\documentclass[aps,prx,twocolumn,superscriptaddress]{revtex4-2}
\usepackage[english]{babel}
\usepackage{amssymb}
\usepackage{amsmath}
\usepackage{mathtools}
\usepackage{txfonts}
\usepackage{mathdots}
\usepackage{array}
\usepackage{amssymb}
\usepackage{amsfonts}
\usepackage{amsmath}
\usepackage{tikz}
\usetikzlibrary{arrows.meta,positioning,calc,decorations.pathreplacing}
\usepackage{standalone}
\colorlet{cFrozen}{black!45}
\colorlet{cActive}{blue!55}
\colorlet{cUpdated}{green!55!black}
\colorlet{cOrd}{orange!85!black}
\colorlet{cDis}{teal!80!black}
\colorlet{cFlow}{blue!45!black}      %
\colorlet{cWeight}{red!55!black}     %

\usepackage{physics}
\usepackage{mathrsfs}
\usepackage{booktabs}
\usepackage{threeparttable}
\usepackage{multirow}
\usepackage{graphicx}
\usepackage{tabularx}
\usepackage{adjustbox}
\usepackage{comment}
\usepackage{etoolbox}
\AtBeginEnvironment{thebibliography}{\interlinepenalty=10000 }  

\usepackage{hyperref}
\usepackage[normalem]{ulem}
\hypersetup{
    unicode=true,     
    pdftoolbar=false,  
    pdfmenubar=true,   
    pdffitwindow=false, 
    pdfstartview={FitH},
    pdftitle={},    
    pdfauthor={Authors},     
    pdfsubject={},   
    pdfcreator={},   
    pdfproducer={}, 
    pdfkeywords={quantum state transfer} {superconducting processor} {quantum state tomography}, 
    pdfnewwindow=true,
    colorlinks=true,
    linkcolor=black,
    citecolor=blue, 
    filecolor=magenta,
    urlcolor=blue,
    psdextra=true,
}
\usepackage[nameinlink]{cleveref}

\crefname{equation}{Eq.}{Eqs.}
\Crefname{equation}{Equation}{Equations}

\crefname{figure}{Fig.}{Figs.}
\Crefname{figure}{Figure}{Figures}
\crefname{table}{Table}{Tables}
\Crefname{table}{Table}{Tables}

\crefname{section}{Sec.}{Secs.}
\Crefname{section}{Section}{Sections}
\crefname{subsection}{Sec.}{Secs.}
\Crefname{subsection}{Section}{Sections}
\crefname{subsubsection}{Sec.}{Secs.}
\Crefname{subsubsection}{Section}{Sections}

\crefname{appendix}{Appendix}{Appendices}
\Crefname{appendix}{Appendix}{Appendices}
\crefname{subappendix}{Appendix}{Appendices}
\Crefname{subappendix}{Appendix}{Appendices}

\creflabelformat{equation}{#2#1#3}

\DeclareMathOperator\supp{supp}
\newcommand{\bookmarkphi}[1]{\texorpdfstring{$\phi^{#1}$}{phi\textasciicircum #1}}

\begin{document}

\title{The Blume{--}Capel Model as a \bookmarkphi{6} Field Theory:
A Monte Carlo and Flow Model Study}
\author{Ejaaz Merali}
\email{emerali@ucdavis.edu}
\affiliation{Department of Physics and Astronomy, University of California, Davis, CA 95616, USA}
\affiliation{Department of Physics and Astronomy, San Jos\'{e} State University, San Jos\'{e}, CA 95192 USA}
\author{Ryan Larson}
\email{rmlarson@ucdavis.edu}
\affiliation{Department of Physics and Astronomy, University of California, Davis, CA 95616, USA}
\author{Ehsan~Khatami}
\email{ehsan.khatami@sjsu.edu}
\affiliation{Department of Physics and Astronomy, San Jos\'{e} State University, San Jos\'{e}, CA 95192 USA}
\author{Richard Scalettar}
\email{scalettar@physics.ucdavis.edu}
\affiliation{Department of Physics and Astronomy, University of California, Davis, CA 95616, USA}

\begin{abstract}
We report Monte Carlo simulations of a 
continuous spin variant of the Blume{--}Capel model.
We compute the phase diagrams in the crystal-field-temperature plane, as a function of the softness of the spins.
We show that normalizing flows, a class of generative models capable of direct sampling without autocorrelation, are effective samplers across the phase diagram, including in the first-order transition region and in the vicinity of the tricritical point.
Finite-size scaling analyses of flow samples locate the first- and second-order transitions as well as the tricritical point with accuracy comparable to canonical and extended-ensemble Monte Carlo.
\end{abstract}

\maketitle

\section{Introduction}

Discrete spin models provide canonical settings for studying phase transitions and for 
developing Monte Carlo algorithms. The Ising model, with $S_i=\pm1$, exhibits a continuous 
thermal transition in dimension $D>1$, and has played a central role in the concrete understanding
of spontaneous symmetry breaking~\cite{Onsager1944Ising,Brush1967Ising,Binder2001MonteCarlo,Pelissetto2002Critical}.  
At the same time, its analysis with computational methods has led to
key developments in the evolution of numerical techniques such as finite-size scaling~\cite{Fisher1972FiniteSize,Cardy1984FiniteSizeScaling,BloteCardyNightingale1986}, Binder cumulants~\cite{Binder1981}, 
and cluster algorithms~\cite{Swendsen1987Cluster,Wolff1989Cluster}. 

The spin-1 Blume–Capel model~\cite{blume1966,capel1966,mukamel1974ising} extends the local state space to $S_i=0,\pm1$ by introducing a 
crystal-field term that controls the population of the $S_i=0$ ``vacancy" state. 
Its two-dimensional phase diagram contains both continuous and first-order transition lines,
meeting at a tricritical point.

The Ising model is well known to have a soft-spin counterpart, the scalar lattice $\phi^4$ model. 
Analogously, a scalar $\phi^6$ theory with three preferred local values 
provides a natural soft-spin counterpart of the Blume–Capel model
~\cite{Moueddene2024BlumeCapel3D,Kwak2015BlumeCapel,Gunaratnam2024Tricritical,Sarbach1979}. 
It also substantially extends Ising phenomenology:
The scalar $\phi^6$ theory is the canonical Landau-Ginzburg description of tricriticality: 
while a positive quartic coupling produces an ordinary continuous transition, 
a negative quartic coupling permits a first-order transition stabilized by the sixth-order term, 
with the two regimes meeting at a tricritical point. In two dimensions, Monte Carlo studies of 
scalar $\phi^6$ theory have confirmed the tricritical-Ising critical behavior 
expected from conformal-field-theory arguments~\cite{Asorey1995Tricritical}.

More recently, a different method based on ideas in machine learning, called normalizing flows, has emerged for the simulation of lattice field theories~\cite{Albergo2019FlowBased,Albergo2021Introduction,DelDebbio2021TrivializingMaps,Gerdes2023EquivariantFlows,Singha2022ConditionalFlow}.
By mapping configurations from a simple, easy-to-sample distribution through a series of invertible transformations, flows can represent complicated, highly correlated distributions.
A well-trained flow can generate independent configurations, which can be used either directly with importance weights or as proposals to a Metropolis sampler thereby dramatically reducing the autocorrelation associated with local or molecular-dynamics-based Markov chains.

So far, most scalar-field benchmarks of normalizing flows have focused on $\phi^4$ theories 
near continuous critical points, where the principal obstacle is critical slowing down.
First-order transitions pose a qualitatively different challenge: the target distribution 
becomes strongly multimodal, with competing phases separated by an exponentially suppressed 
interfacial region. This makes mode collapse and rare-event sampling potential challenges for flow 
models~\cite{Bai2024FlowFirstOrder}.

A $\phi^6$ realization of Blume–Capel physics therefore provides a controlled setting in which the 
character of the sampling problem can be tuned continuously from ordinary criticality through 
tricriticality into a first-order coexistence regime.

With that motivation in mind, in this work we use a two-dimensional scalar $\phi^6$ model to study both tricritical statistical mechanics and the performance of flow-based Monte Carlo simulations. 
The onsite potential continuously interpolates between a soft scalar field and the three-state structure of the Blume–Capel model, allowing us to determine how the phase boundary and tricritical point evolve as the local field becomes increasingly discrete. 
 
More importantly, the same model provides an algorithmic testbed in which the dominant sampling difficulty changes qualitatively: from critical slowing down along the continuous transition line, to tricritical fluctuations, and finally to phase coexistence and free-energy barriers along the first-order line. 
We benchmark normalizing-flow samplers against cluster and extended-ensemble Monte Carlo throughout these regions. 
A systematic comparison across all three regimes of a single two-dimensional $\phi^6$ phase diagram is, to the best of our knowledge, lacking.

Independently of this algorithmic comparison, we obtain the phase diagrams of the soft-spin Blume–Capel model and clarify their hard-well connection to the discrete model. 
A key ingredient in our analysis is the inclusion of an entropic $T\ln2$ shift arising from the unequal curvature of the vacancy well relative to the two wells corresponding to (occupied) Ising spins.

Our paper is organized as follows:  We first provide, in \cref{sec:soft_spin_models}, a more detailed description of the continuous spin variants of the Ising and Blume{--}Capel models.
We also present results from `conventional' Monte Carlo techniques which give the critical temperature $T_c$ in the Ising case.
\Cref{sec:results} then compares results of conventional Monte Carlo with those of flow model methods.
There are two key outcomes of our investigations.
The first is the determination of the phase diagrams in a way which is especially transparent in their connection to the underlying discrete spin cases, and which has not, to our knowledge, yet been treated in the literature.
The second is a careful study of the efficacy and the advantages of flow model methods in the presence of combined second and first order transition regimes, and hence for situations in which a tricritical point is present.
\Cref{app:mft,app:flow} contain a discussion of the phase diagrams within mean field theory and of the detailed anatomy of the flow model, respectively.

\section{Soft-Spin Ising and Blume{--}Capel Models} \label{sec:soft_spin_models}
We begin with the $\phi^4$ analogue because it provides a control case in which the transition remains 
continuous throughout the interpolation to the hard-spin Ising limit. It therefore separates effects 
associated simply with softening the local spin amplitude from the qualitatively new tricritical 
and first-order behavior introduced by the $\phi^6$ theory.
This section also serves the 
purpose of providing the critical temperature of $\phi^4$ field theory in a way which is connected
in a clear and simple way to its Ising progenitor.  

\subsection{The Soft-Spin Ising Model: \bookmarkphi{4}}\label{ssec:ising_phi4}

The two models in question are,
\begin{eqnarray}
E &=& -J \sum_{\langle ij \rangle} S_i S_j 
\nonumber \\
E &=& -J \sum_{\langle ij \rangle} \phi_i \phi_j 
+ \lambda \sum_i (\phi_i^2-1)^2
\label{eq:Ising}
\end{eqnarray}
where in the top line $S_i= \pm 1$ is discrete while in the bottom line
$-\infty < \phi_i < +\infty$ is continuous.  The stucture of the
potential  $V(\phi_i) = \lambda (\phi_i^2-1)^2$
is such that $\phi_i = \pm 1$ minimizes $V$, and these values become
increasingly favored as $\lambda \rightarrow \infty$.
In both cases the degrees of freedom are 
coupled on near-neighbor sites $\langle ij \rangle$ by an exchange constant $J$.
We will consider the ferromagnetic case $J>0$, although on a bipartite lattice
$J>0$ and $J<0$ can be mapped onto each other via $\phi_{i} \leftrightarrow -\phi_{i}$
on one of the two sublattices.
Our geometry is that of an $L\times L$ square lattice.

If we were to change the scale of the spins $S_i = \pm S_0$ the Ising
transition temperature would trivially scale by the factor $S_0^2$.
To isolate the effect of relaxing the restriction to discrete values in
the continuous model from this effect of the spin `length' on $T_c$ we
can instead report values for $T_c/\ev*{\phi^2}$.
Indeed, as noted in \cref{app:phi4_mft}, within mean field theory this normalized critical temperature of the 
continuous spin model, $T_c/\ev*{\phi^2} = 4J$, is then independent of $\lambda$ on a square lattice.
The continuous and discrete spin models
have identical mean field critical points.  

\begin{figure}[btp!]
    \centering
    \includegraphics[width=\columnwidth]{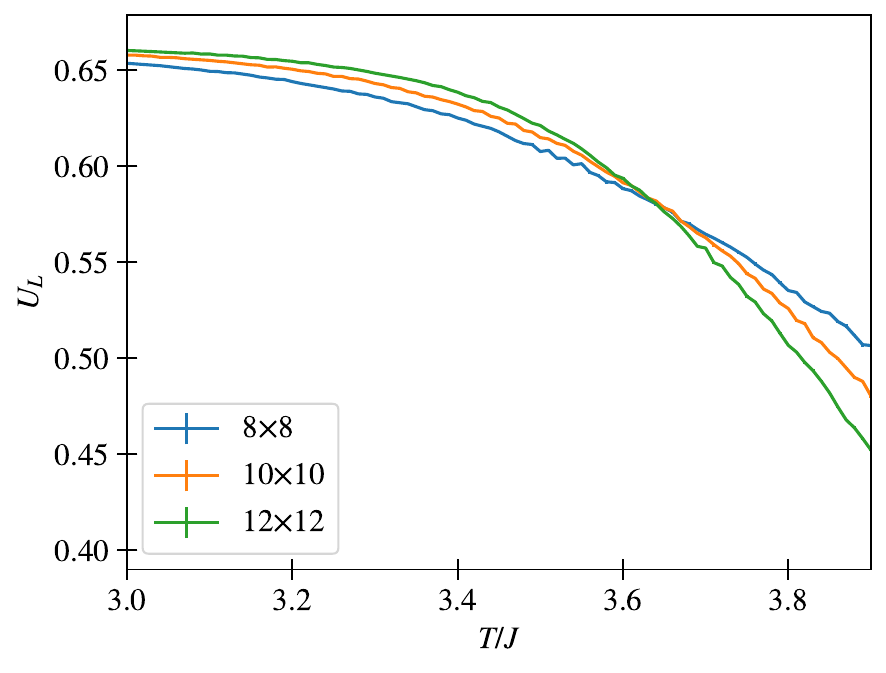}
    \caption{Critical temperatures are found using the crossings of the Binder parameter $U_L$ for different linear system sizes $L$.
    In this representative example for the continuous spin variant of the Ising model [\cref{eq:Ising}]  on the square lattice, we chose 
    $\lambda/J = 0.5$, resulting in a critical temperature at $T_c/J=3.641$.
    }
    \label{fig:phi4Binder}
\end{figure}

We have used conventional single spin flip Metropolis Monte
Carlo to obtain the critical temperature of the continuous spin variant in \cref{eq:Ising}.
Specifically, we evaluated the Binder ratio $U_L = 1 - \ev*{m^4} / 3\ev*{m^2}^2$,
with $m=L^{-2} \sum_i \phi_i $,
as a function of temperature
for different linear lattice sizes $L=8,12,16$ and used the resulting crossing, an example of which is shown in \cref{fig:phi4Binder}, to determine $T_c$~\cite{Binder1981}.

\begin{figure}[btp!]
    \centering
    \includegraphics[width=\columnwidth]{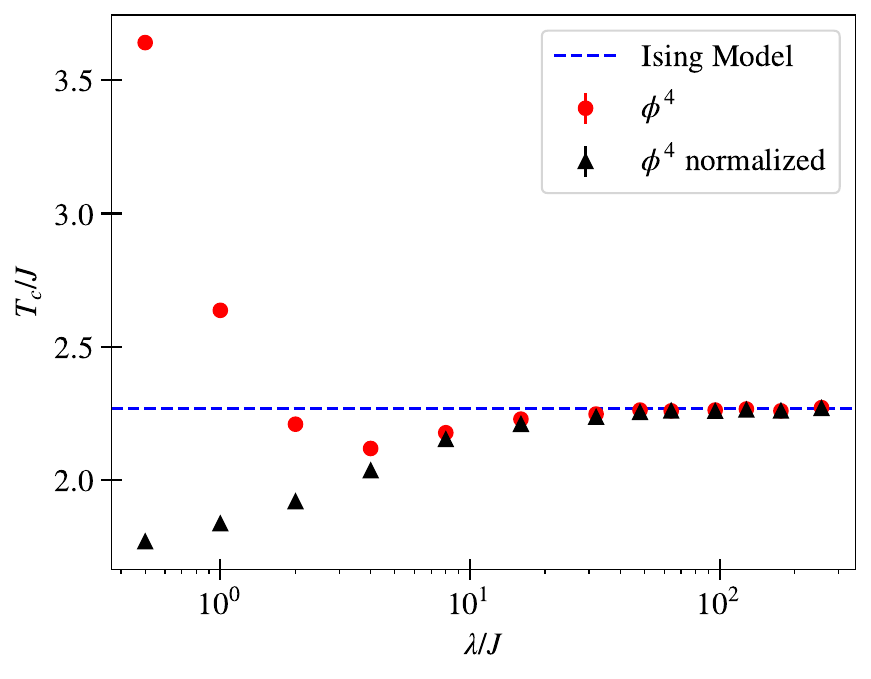}
    \caption{
    Critical temperatures of the continuous spin variant of the Ising model [\cref{eq:Ising}] on the square lattice as a function of the parameter $\lambda$, which tunes the amount the
    degrees of freedom deviate from their discrete values $\pm 1$.
    The limit of large $\lambda$ recovers the Onsager value $T_c/J = 2.269$.
    The red circles are $T_c/J$ while
    the black triangles are normalized by $\langle \phi^2 \rangle$. 
    The value of $T_c$ at the smallest $\lambda/J=0.5$ corresponds to the Binder crossing of \cref{fig:phi4Binder}.
    Where error bars aren't visible they lie inside of the point markers.
    }
    \label{fig:Tcphi4}
\end{figure}

\begin{figure}[t!]
    \centering
    \includegraphics[height=3.0in,width=3.60in]{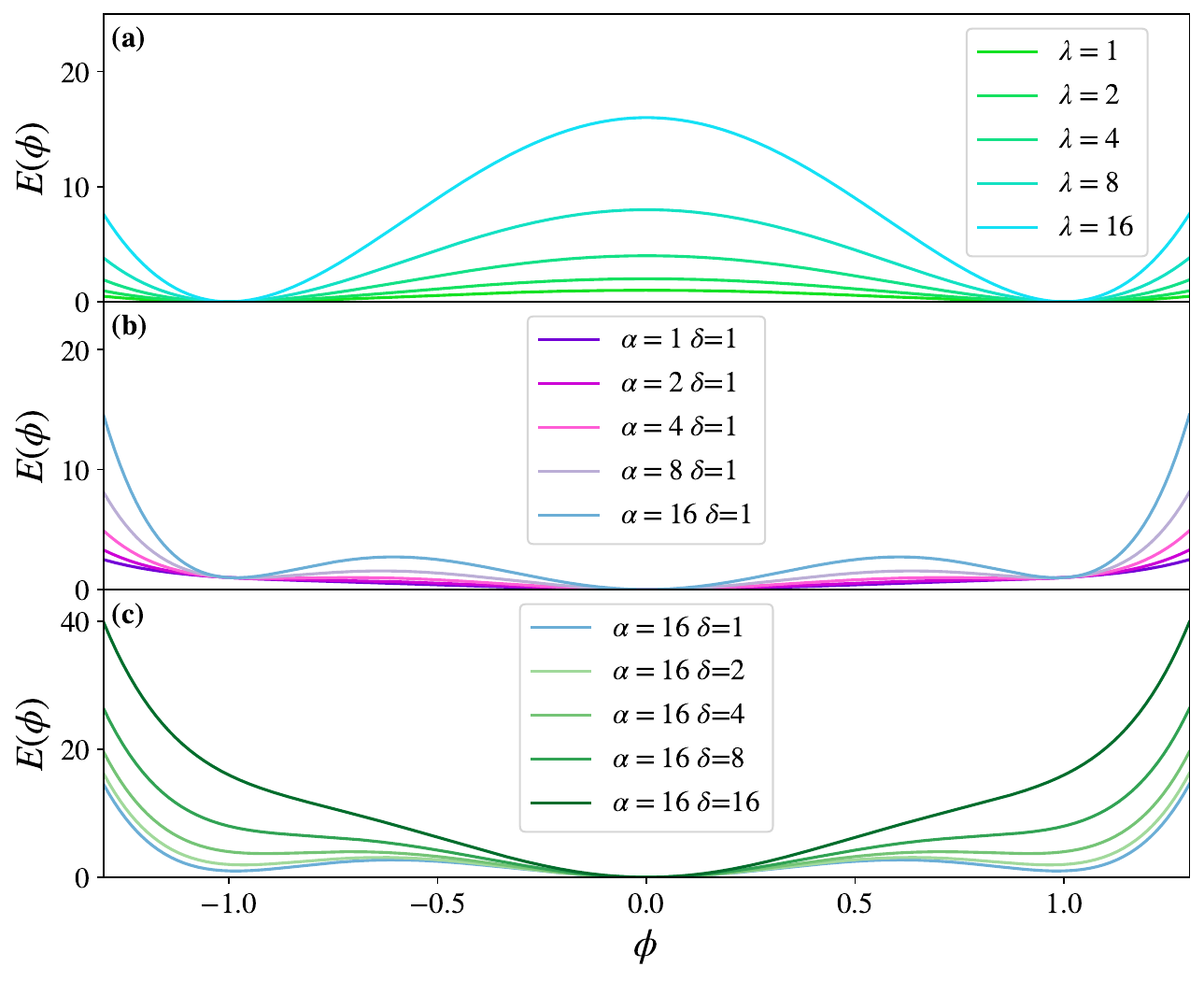}
    \caption{(a) The potential $V(\phi)$ for four values of $\lambda$ in \cref{eq:Ising}, the soft-spin variant of the Ising model. 
    With increasing $\lambda$ the wells at $\phi=\pm 1$ become deeper.
    (b) shows the potential of \cref{eq:phi6}, the soft-spin variant of the Blume{--}Capel model,  for fixed vacancy potential $\delta=1$ and varying $\alpha$, the parameter which
    tends to confine $\phi$ to $\phi=0,\pm1$.  
    (c) shows fixed $\alpha=16$ at several $\delta$.  Increasing $\delta$ favors
    vacancies ($\phi=0$).
    In both (b) and (c), the lower curvature of the $\phi=0$ well relative to the $\phi=\pm1 $ wells is evident.
    This blurs the mapping to the Blume{--}Capel limit.  See text.
    }
    \label{fig:Vphi}
\end{figure}

\Cref{fig:Tcphi4} shows $T_c(\lambda)$ from a collection of such crossings.
Although the un-normalized $T_c$ can exceed the Ising value, the normalized critical
temperature is steadily suppressed as $\lambda$ decreases.  This reflects the expected effect
of  the increased entropy associated with the continuous $\phi$ field over the discrete
Ising case, once the trivial `length' effect is removed.
Panel (a) of \cref{fig:Vphi} plots $V(\phi)$ for four values of $\lambda$,
illustrating the softening of the potential with the lowering of $\lambda$.

\subsection{The Soft-Spin Blume{--}Capel Model: \bookmarkphi{6}}\label{ssec:bc_phi6}

The
Blume{--}Capel model~\cite{blume1966,capel1966}
\begin{eqnarray}
E = -J \sum_{\langle ij \rangle} S_i S_j + \Delta \sum_i S_i^2
\label{eq:BC}
\end{eqnarray}
is a generalization of the Ising model to include vacancies.
Here the spin $S_{i}$ takes three possible values $S_i = \pm 1,0$, 
with $S_{i}=0$ modeling the absence of a spin on site $i$.
The \emph{crystal field} $\Delta$ controls the occupation density, penalizing spins taking non-zero values.
As with the Ising model, the sum in the first term is usually taken to be over near-neighbor sites $\langle ij \rangle$ of a lattice.
For  $J<0$, \cref{eq:BC} is relevant to  $^{3}$He-$^{4}$He mixtures~\cite{blume1971,graf1967phase}, while $J>0$ is appropriate for binary alloys~\cite{lajzerowicz1975spin,mukamel1974ising} and superionic conductors~\cite{girvin1977cation}.

On a square lattice, the phase diagram of the Blume{--}Capel model in the $(T,\Delta)$ plane has been studied with Monte Carlo
simulations~\cite{wang1987phase,care1993microcanonical,hurt2006analysis}, finite-size
scaling~\cite{beale1986finite,xavier1998critical}, 
renormalization group
methods~\cite{berker1976blume,burkhardt1976application}, and series
expansions~\cite{saul1974tricritical}.  
One of its interesting features is that the phase transition between magnetically ordered and disordered phases
is initially second order for $\Delta \rightarrow -\infty$, the Ising limit,
but that as $\Delta$ increases there is a tricritical point at which the transition becomes first order.
The position of this tricritical point is known to high accuracy: 
$(T_t/J,\Delta_t/J) = (0.609,1.966)$ ~\cite{beale1986finite,silva2006}.

In analogy with the Ising case,
we can consider a continuous spin generalization of 
\cref{eq:BC},
\begin{align}
E = J \sum_{\langle ij \rangle} \phi_{i} \, \phi_{j} + \alpha \sum_i \phi_{i}^2 (\phi_{i}^2 - 1)^2
 + \delta \sum_i \phi_{i}^2 
\label{eq:phi6}
\end{align}
In this case 
the parameter $\alpha$ controls the preference of  the continuous $\phi_{i}$ to take on the values
$\phi_{i} = \pm1, 0$ since its energy is minimized there.
We will often denote the full single-site potential as $V=V_\alpha+V_\delta$.
$V_\delta$ connects \cref{eq:phi6} directly to \cref{eq:BC}.

It is interesting to note that unlike the mapping between the Ising model and 
$\phi^{4}$ field theory, where one can approach the Ising limit precisely at large
$\lambda$, \cref{eq:BC} is not obtained exactly from
\cref{eq:phi6} as $\alpha \rightarrow \infty$.
The reason is that even though the potential takes the same (zero) value
for $\phi=0$ as for $\phi=\pm1$, the curvature is different.  Expanding
to second order about the minima at $\phi_{i} = \pm 1$ gives a coefficient
$d^{2} V_{\alpha} / d \phi^{2}=8 \alpha$
while the curvature 
$d^{2} V_{\alpha} / d \phi^{2}=2 \alpha$
is softer at $\phi=0$.

The effect of this difference in curvature can be made more precise.
In the large-$\alpha$ limit, the Gaussian weight of a well is proportional to $1/\sqrt{d^2V/d\phi^2}$. Therefore,
the weights of the unoccupied ($\phi_i=0$) and occupied ($\phi_i=\pm 1$) minima obey
$ w_{\pm}/w_0 = e^{-\beta\delta}/2.$
For the discrete Blume–Capel model the corresponding weight of each \(S=\pm1\) state relative to \(S=0\) is
$ e^{-\beta \Delta}$.
Thus the soft model approaches a discrete Blume–Capel model with
$
\Delta
=
\delta+T\ln2.$
Equivalently,
$\delta=\Delta-T\ln2,$
yielding a parameter-free prediction for the $\alpha \to \infty$ phase boundary:
$ \delta_c(T) = \Delta_c(T)-T\ln2.$
At the known discrete tricritical point,
the corresponding hard-well limit of $\delta$ is approximately
$ \delta_t/J \simeq 1.966-0.609\ln2 \simeq 1.544$,
making use of the well-established discrete tricritical values \cite{silva2006}.

\section{Monte Carlo and Flow Model Results for \bookmarkphi{6} Field Theory} \label{sec:results}

Here we contrast the effectiveness of conventional Monte Carlo methodology
and flow model approaches for obtaining the phase diagram of \cref{eq:phi6}.  
Unlike the simple single spin flip technique employed in \cref{ssec:ising_phi4} for \cref{eq:Ising},
we employ here the more sophisticated and powerful ideas of Wolff updates and parallel
tempering to provide an appropriate challenge to the flow model approach.

\begin{figure}[btp!]
    \centering
    \includegraphics[width=\columnwidth]{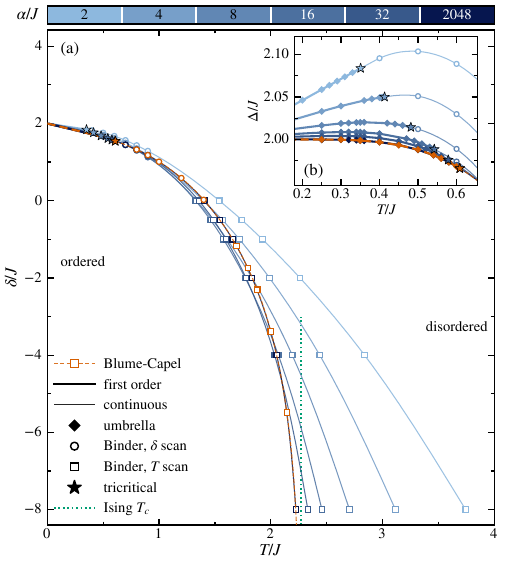}
    \caption{Phase diagram of the $\phi^6$ theory of \cref{eq:phi6}. The inset zooms in on the region near the tricritical points, where we've plotted against the Blume{--}Capel crystal field $\Delta$ in order to spread out the different $\alpha$ curves for legibility.
    Lines for the continuous transition are provided as guides to the eye, while the first order curve is actually made of many densely packed measurements, with only a few markers displayed for legibility. The vertical dashed line is the Ising $T_c/J \simeq 2.269$ to which the critical temperature
    of the discrete spin Blume{--}Capel model converges as $\delta \rightarrow -\infty$. Error bars are smaller than the markers.}
    \label{fig:phi6_bc_phase_diagram}
\end{figure}

\begin{figure}[btp!]
    \centering
    \includegraphics[width=\columnwidth]{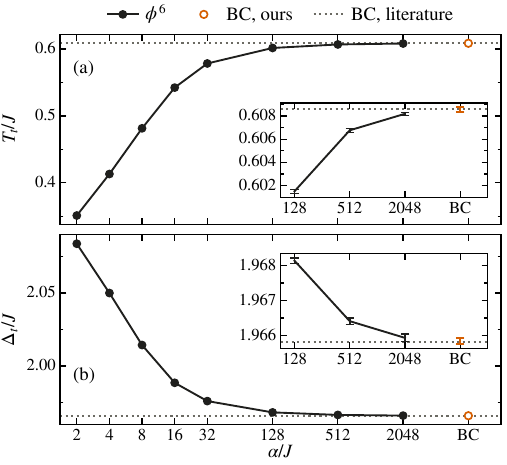}
    \caption{Convergence of the position of the soft-spin tricritical point $T_t/J$ (panel (a)) and $\Delta_t/J$ (panel (b)) to the discrete Blume{--}Capel (BC) limits.
    Error bars are smaller than the markers of the zoomed out plots and are only displayed in the inset.
    Literature value is from \cite{Jung_2017}.}
    \label{fig:tricritical_alpha}
\end{figure}

\subsection{`Advanced' Markov Chain Monte Carlo}

The $\phi^6$ and Blume{--}Capel models were sampled via Markov-chain Monte Carlo (MCMC), using a combination of single-site updates and Wolff cluster updates of embedded Ising signs~\cite{Wolff1989Cluster,Brower1989}. 
The single-site updates involved generating proposals from a three component Gaussian mixture fit to the single-site potential $V(\phi)$, and accepting with a Metropolis-Hastings step~\cite{metropolis1953,hastings1970}.
This allows efficient switching between wells along with exploration within each.
We make use of the bipartite structure of the lattice to vectorize our single-site updates by updating all spins within a sublattice in parallel~\cite{Wansleben1984MonteCS,Landau_Binder_2021}.

Although Wolff updates are efficient near the second-order critical line, they face difficulty when approaching the tricritical point, where the cluster update's dynamical critical exponent degrades to the local update value~\cite{jeon2026clusterdynamicsstayfastuntil}.
Near the first-order transition, the coexistence of two phases separated by a large energy barrier poses difficulties for MCMC algorithms, which cannot be overcome by the Wolff algorithm either.

Thus, we make use of the replica exchange method~\cite{geyer_replica1991,swendsen_wang_replica1986,hukushima_replica1996}, using swaps in $\delta$ since swaps between temperature rungs alone rarely make enough round trips as the system size grows.
As an additional cross-check and for parameter points closer to the tricritical point from the first-order side, we make use of umbrella sampling~\cite{torrie1977} by applying a harmonic bias in the density $\rho = L^{-2}\sum_i\phi_i^2$ to the energy.
Several independent simulations are run with the harmonic biases pinning the density to a grid of overlapping windows, which are then combined using the weighted-histogram analysis method~\cite{ferrenberg1989,kumar_wham1992}, yielding a histogram $P(\rho)$.
Reweighting the histogram in $\delta$ we can find the point of minimum density between the two histogram peaks. The critical field for the finite-size system is then taken to be that $\delta_c$ which makes the integral of the two peaks (separated by the cut) equal after accounting for the degeneracy of the ordered states.
Close to the tricritical point, the two peaks begin to merge.

\begin{figure*}[btp!]
    \centering
    \providecommand{\mscoarselabel}{coarse blocks}%
\providecommand{\msfinelabel}{fine blocks}%
\def\bw{58}%
\def\dg{8}%
\def\mw{18}%
\pgfmathsetmacro{\blw}{(\bw-3.2-\dg)/2}%
\pgfmathsetmacro{\cm}{(\blw-18)/3}%
\def\xcl{54}%
\pgfmathsetmacro{\xcr}{\xcl+\bw}%
\pgfmathsetmacro{\xpc}{\xcr+12}%
\pgfmathsetmacro{\xml}{\xpc+33}%
\pgfmathsetmacro{\xmc}{\xml+\mw/2}%
\pgfmathsetmacro{\xmr}{\xml+\mw}%
\pgfmathsetmacro{\xfm}{\xmr+13}%
\pgfmathsetmacro{\xfl}{\xfm+13}%
\pgfmathsetmacro{\xfr}{\xfl+\bw}%
\pgfmathsetmacro{\xph}{\xfr+13}%
\pgfmathsetmacro{\xtot}{\xph+6}%
\providecommand{\multiscaletextscale}{1}%
\ifstandalone\providecommand{\multiscalewidth}{179mm}\else\providecommand{\multiscalewidth}{\linewidth}\fi%
\pgfmathsetlengthmacro{\msunit}{\multiscalewidth/(\xtot+0.4)}%
\pgfmathsetlengthmacro{\mselbow}{2.5*\msunit}%
\pgfmathsetlengthmacro{\msarrow}{2.6*\msunit}%
\pgfmathsetlengthmacro{\mscorner}{0.9*\msunit}%
\setbox0\hbox{\normalsize N}\pgfmathsetlengthmacro{\mscap}{0.5*\multiscaletextscale*\the\ht0}%
\begin{tikzpicture}[
  x=\msunit,y=\msunit,
  font=\normalsize,
  >={Stealth[length=\msarrow]},
  lbl/.style={font=\normalsize,align=center,scale=\multiscaletextscale},   %
  px1/.style={line width=0.24pt},
  px2/.style={line width=0.48pt},
  px3/.style={line width=0.72pt},
  stack/.style={rounded corners=\mscorner,px2,fill=cFlow!10,draw=cFlow!45},     %
  fixed/.style={rounded corners=\mscorner,px2,fill=black!4,draw=black!40},      %
  tab/.style={lbl,fill=white,draw=cFlow!45,px2,rounded corners=\mscorner,inner xsep=4.5pt,inner ysep=3.5pt,anchor=west},
]
\renewcommand{\baselinestretch}{1}   %
\pgfmathsetmacro{\ycnn}{1.75}   %
\def\gap{1.0}                  %
\def\up{16}                    %
\colorlet{fillBase}{cDis!14}   \colorlet{inkBase}{cDis}        \colorlet{lineBase}{cDis!60}
\colorlet{fillFlow}{cUpdated!18}\colorlet{inkFlow}{cUpdated}   \colorlet{lineFlow}{cUpdated!70}
\colorlet{latFrozen}{cFrozen!35}
\colorlet{latActive}{cActive!80}

\newcommand{\fine}[4]{%
  \begin{scope}[shift={(#1-6,#2-6)}]
    \fill[#3] (0,0) rectangle (12,12);
    \draw[#4!55,px1,step=3] (0,0) grid (12,12);
    \draw[#4,px2,step=6] (0,0) grid (12,12);
  \end{scope}}
\newcommand{\coarse}[4]{%
  \begin{scope}[shift={(#1-3,#2-3)}]
    \fill[#3] (0,0) rectangle (6,6);
    \draw[#4,px2,step=3] (0,0) grid (6,6);
  \end{scope}}
\newcommand{\detail}[4]{%
  \foreach \dk in {2,1,0}{
    \begin{scope}[shift={(#1-3.7+\dk*0.7,#2-3.7+\dk*0.7)}]
      \fill[#3] (0,0) rectangle (6,6);
      \draw[#4,px2,step=3] (0,0) grid (6,6);
    \end{scope}}}

\newcommand{\couplinglayer}[4]{%
  \draw[rounded corners=\mscorner,px2,fill=white,draw=cFlow!60] (#1-4.5,#2-6.5) rectangle (#1+4.5,#2+6.5);
  \begin{scope}[shift={(#1-3,#2-3)}]
    \pgfmathsetmacro{\cs}{6/#4}
    \pgfmathtruncatemacro{\nlast}{#4-1}
    \fill[latFrozen] (0,0) rectangle (6,6);
    \foreach \ci in {0,...,\nlast}{\foreach \cj in {0,...,\nlast}{
      \pgfmathparse{int(mod(\ci+\cj+#3,2))}
      \ifnum\pgfmathresult=1 \fill[latActive] (\ci*\cs,\cj*\cs) rectangle ++(\cs,\cs);\fi}}
    \draw[black!40,px1,step=\cs] (0,0) grid (6,6);
  \end{scope}}
\newcommand{\block}[3]{%
  \draw[rounded corners=\mscorner,fill=cFlow!4,draw=cFlow!45,px2] (#1,#2-8.1) rectangle (#1+\blw,#2+8.1);
  \couplinglayer{#1+\cm+4.5}{#2}{0}{#3}
  \couplinglayer{#1+\blw-\cm-4.5}{#2}{1}{#3}}
\newcommand{\couplingstack}[4]{%
  \draw[stack] (#1,#2-9.9) rectangle (#1+\bw,#2+12.6);
  \node[tab] at (#1+3,#2+12.6) {#3};
  \block{#1+1.6}{#2}{#4}
  \node[lbl] at (#1+\bw/2,#2) {$\cdots$};
  \block{#1+\bw-1.6-\blw}{#2}{#4}}

\fine{6}{0}{fillBase}{lineBase}
\node[lbl,below] at (6,-6) {$z$};
\draw[fixed] (19,-4.5) rectangle (33,4.5);  \node[lbl,anchor=mid] at (26,0) {split};
\draw[->,inkBase,px3] (12+\gap,0) -- (19-\gap,0);

\coarse{44}{\up}{fillBase}{lineBase}
\node[lbl,above] at (44,\up+3) {$z_c$};
\detail{44.7}{-\up}{fillBase}{lineBase}
\node[lbl,below] at (44.7,-\up-3.7) {$r$};
\draw[->,inkBase,px3,rounded corners=\mselbow] (26,4.5+\gap) |- (41-\gap,\up);
\draw[->,inkBase,px3,rounded corners=\mselbow] (26,-4.5-\gap) |- (41-\gap,-\up);

\couplingstack{\xcl}{\up}{\mscoarselabel}{2}
\draw[->,inkBase,px3] (47+\gap,\up) -- (\xcl-\gap,\up);
\coarse{\xpc}{\up}{fillFlow}{lineFlow}
\node[lbl,above] at (\xpc,\up+3) {$z_c^\prime$};
\draw[->,inkFlow,px3] (\xcr+\gap,\up) -- (\xpc-3-\gap,\up);

\draw[stack] (\xpc-16,-\up-6.5) rectangle (\xpc+16,-\up+6.5);
\node[lbl] at (\xpc,-\up) {$\zeta=e^{s}r+t$};
\draw[->,inkBase,px3] (48.8+\gap,-\up) -- (\xpc-16-\gap,-\up);
\draw[stack] (\xpc-7.5,\ycnn-3.75) rectangle (\xpc+7.5,\ycnn+3.75);
\node[lbl,anchor=base,yshift=-\mscap] at (\xpc,\ycnn) {CNN};
\draw[->,inkFlow,px3] (\xpc,\up-3-\gap) -- (\xpc,\ycnn+3.75+\gap);
\draw[->,px2] (\xpc,\ycnn-3.75-\gap) -- node[lbl,right,inner sep=3pt] {$s,\,t$} (\xpc,-\up+6.5+\gap);
\detail{\xpc+25.7}{-\up}{fillFlow}{lineFlow}
\node[lbl,below] at (\xpc+25.7,-\up-3.7) {$\zeta$};
\draw[->,inkFlow,px3] (\xpc+16+\gap,-\up) -- (\xpc+22-\gap,-\up);

\draw[fixed] (\xml,-4.5) rectangle (\xmr,4.5);  \node[lbl,anchor=mid] at (\xmc,0) {merge};
\draw[->,inkFlow,px3,rounded corners=\mselbow] (\xpc+3+\gap,\up) -| (\xmc,4.5+\gap);
\draw[->,inkFlow,px3,rounded corners=\mselbow] (\xpc+29.8+\gap,-\up) -| (\xmc,-4.5-\gap);
\fine{\xfm}{0}{fillFlow}{lineFlow}
\draw[->,inkFlow,px3] (\xmr+\gap,0) -- (\xfm-6-\gap,0);

\couplingstack{\xfl}{0}{\msfinelabel}{4}
\draw[->,inkFlow,px3] (\xfm+6+\gap,0) -- (\xfl-\gap,0);
\fine{\xph}{0}{fillFlow}{lineFlow}
\node[lbl,below] at (\xph,-6) {$\phi$};
\draw[->,inkFlow,px3] (\xfr+\gap,0) -- (\xph-6-\gap,0);

\end{tikzpicture}
    \caption{
        The overall structure of the flow model. A configuration $z$ is drawn from an $L\times L$ base density and then each $2\times2$ block is split into a mean $z_c$ and detail coefficients $r$ via a Haar transform. The block-mean is then passed through a sequence of coupling layers, giving $z_c^\prime$ which are then passed through a small convolutional neural network (CNN) to produce parameters $s, t$ that are applied to the details via an affine transformation giving $\zeta$. The $z_c^\prime$ and the updated detail coefficients $\zeta$ are then merged by an inverse Haar transform giving a configuration of size $L\times L$. This is then passed through another sequence of coupling layers, finally giving the field configurations $\phi$.
    }
    \label{fig:flow_model}
\end{figure*}
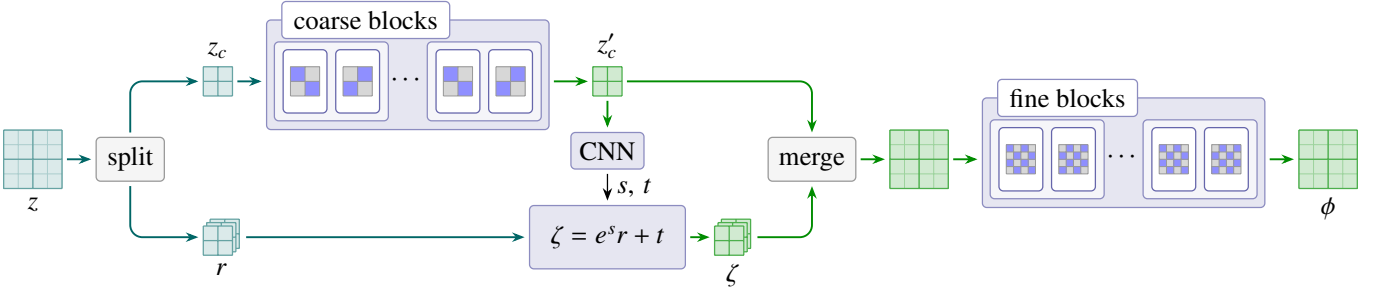

The tricritical point for the finite size system is found by first locating (again through reweighting) the $\delta$ which makes the skewness
\begin{equation}
    \mathcal{S}_\rho = \frac{\ev{(\rho - \ev{\rho})^3}}{\ev{(\rho - \ev{\rho})^2}^{3/2}}
\end{equation}
of the density histogram zero.
The Binder cumulant for the (centered) density is then computed on the zero skew curve, and the tricritical temperature $T_t$ (and hence also $\delta_t$) are extracted via fitting to the scaling form
\begin{equation}\label{eq:density_binder_scaling}
    U_\rho = U_\rho^\ast + a_1 x + a_2 x^2 + \frac{b}{L}(1 + r a_1 x)
\end{equation}
with $x = L^{y_g}(T - T_t)$, $y_g = 4/5$ the second thermal exponent of the tricritical Ising universality class~\cite{den_nijs_1979,nienhuis_1979,friedan_1984,Kwak2015BlumeCapel}, and $U_\rho^\ast, a_1, a_2, b, r$ are parameters that we fit.
We perform our fits across different $\alpha$ using common $U_\rho^\ast$ and $r$ as they are known to be universal~\cite{Wilding_1996}.

The full phase boundaries for several values of the pinning-field $\alpha$ along with the Blume{--}Capel case are shown in \cref{fig:phi6_bc_phase_diagram}.
{\it This figure shows the central result of the first portion of this paper.}  Although the connection between the discrete and continuous
spin models of \cref{eq:BC,eq:phi6} has been noted previously, especially within the context of tricritical behavior
~\cite{Moueddene2024BlumeCapel3D,Kwak2015BlumeCapel,Gunaratnam2024Tricritical,Sarbach1979,Asorey1995Tricritical},
the phase boundaries for different values of the parameter $\alpha$ controlling the depth of the wells at $\phi=-1,0,+1$ have not
been explicitly obtained.  It is noteworthy that the critical $\delta/J$ at low temperatures is
independent of $T/J$, and that the ordered phase extends to higher $T$ with decreasing $\alpha$.
Several of these features emerge from a mean field analysis, as discussed in \cref{app:mft}.

We additionally show the $\alpha\to\infty$ behavior of the tricritical point in \cref{fig:tricritical_alpha}, confirming its convergence to the Blume{--}Capel limit.
$\Delta_t/J$ shows little dependence on $\alpha$, varying only by a few percent as $\alpha$ changes by three orders of magnitude.
This is likely a consequence of the topology of the BC phase boundary, where $\Delta$ is almost $T$ independent in the first order regime.
The tricritical temperature $T_t/J$, on the other hand, increases by almost a factor of two as the spins harden.

\subsection{Flow Model}

In a flow model, one parametrizes a smooth invertible transformation $f:\mathbb{R}^N \to \mathbb{R}^N$ from one distribution over continuous random variables to another.
We call the input distribution the \textit{base density} $p_b(z)$, and compute samples $\phi = f(z)$, where $z \sim p_b(z)$.
The density over $\phi$ is then given by the standard change-of-variables formula for continuous random variables:
\begin{equation}\label{eq:flow_change_of_variables}
    p_\theta(\phi) = p_b(f^{-1}(\phi))\abs{\det \pdv{f^{-1}(\phi)}{\phi}}.
\end{equation}
Our goal is then to construct a function $f$ which maps an easy to sample distribution $p_b$ to a more complex one, in our case the $\phi^6$ lattice field theory $p(\phi) = e^{-\beta E(\phi)}/Z$.
The main design constraints in constructing the function $f$ are two-fold: it must be easily invertible and its Jacobian must have an efficiently computable determinant, which can be accomplished by requiring that the Jacobian be triangular.
This is achieved by using coupling layers~\cite{dinh2017density} in which the variables are split into two groups and each group is transformed conditioned on the values of the other group. 
The conditioning transformation is implemented by rational quadratic splines~\cite{durkan2019neuralsplineflows} whose parameters are produced by a convolutional neural network.
In order to more efficiently model long range behavior, we adopt a multiscale architecture, in which the initial parts of the model apply transformations only to coarse degrees of freedom before applying further transformations to finer ones~\cite{yu2020waveletflowfasttraining, wavelet_crg}. 
A high-level diagrammatic overview of our model is shown in \cref{fig:flow_model}.
All symmetries of the lattice field theory are imposed on the model to aid in the convergence.
Our model is trained without data by minimizing the (reverse) Kullback-Liebler (KL) divergence, equivalent to minimizing the free energy of the model with respect to the target energy.
We refer readers to \cref{app:flow} for more details on our model architecture, implementation, and training procedure.

\begin{figure}[btp!]
    \centering
    \includegraphics[width=\linewidth]{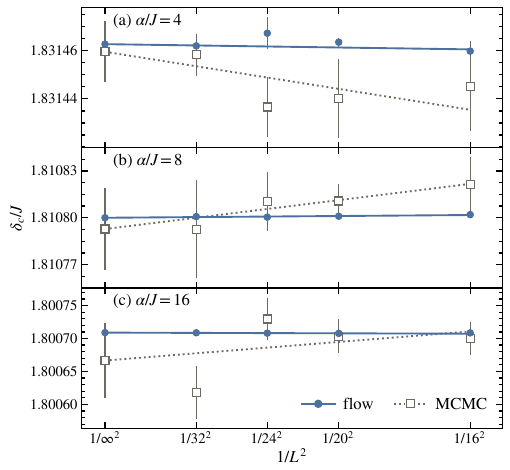}
    \caption{$\delta_c$ for the first order transition through the $T=0.3J$ slice. Extrapolation to the infinite size limit for the flows gives an almost flat curve as expected. MCMC on the other hand is dominated by noise at these scales, resulting in a steeper line.}
    \label{fig:flow_first_order}
\end{figure}

To assess the capabilities of normalizing flows to effectively model the various parts of the phase boundary, we use them to find the locations of the phase transitions along certain parameter slices.
In all experiments described below, we train the flows with 3 different seeds and average the results.
We begin with the slice $T=0.3$ and attempt to extract the $\delta_c$ value corresponding to the first order transition for several $\alpha$, using system sizes $L=16,20,24,32$.
Similarly to the MCMC case, we do this by finding the $\delta$ for which the weight of the ordered phase is twice that of the disordered phase, where the factor of two comes from the degeneracy of the ordered phase.
As described in \cref{ssec:walled_flows}, we use two separate flows near first order transitions, each restricted to one of the two phases.
Because the flows are normalized, we can estimate the partition function restricted to the phase covered by each flow, and the $\delta_c(L)$ is the value for which $Z_O(\delta) = 2\,Z_D(\delta)$.
We perform reweighting in $\delta$ in order to find the root of this equation.
The expected finite size scaling form for $\delta_c(L)$, to leading order, is
\begin{equation}\label{eq:fss_first_order}
    \delta_c(L) = \delta_c^{\infty} + \frac{a}{L^2}
\end{equation}
and we expect the parameter $a$ to be zero due to our condition that $\delta_c(L)$ be the point at which $Z_O = 2\,Z_D$~\cite{janke1992,janke1993}.
We illustrate the flow pair's ability to properly detect $\delta_c$ in \cref{fig:flow_first_order}, and we extract the infinite size limit by fitting \cref{eq:fss_first_order} with the parameter $a$ left free.
The flow pairs are able to give remarkably consistent estimates for the critical $\delta$, resulting in fits that give almost horizontal lines on the scales of the plots.
In contrast, the MCMC umbrella sampler's statistical fluctuations are larger, giving a relatively less precise fit.

\begin{figure*}
    \centering
    \includegraphics[width=\linewidth]{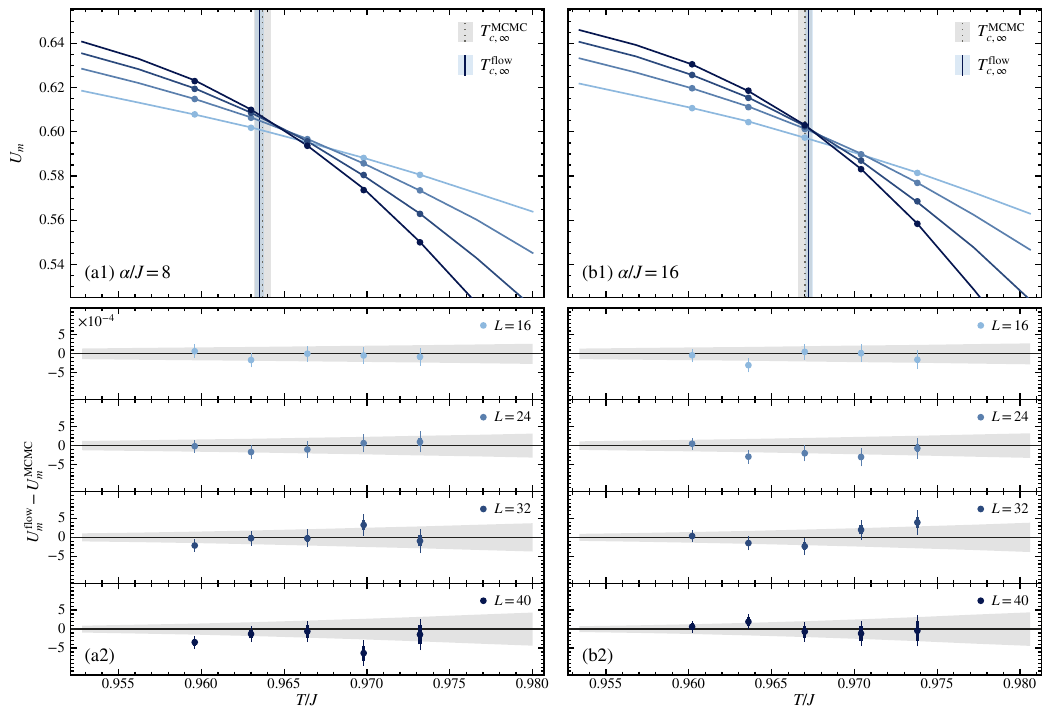}
    \caption{Upper panels: Binder cumulants of the magnetization taken along the $\delta = 1$ slice through the second order phase transition estimated by flow models (error bars are smaller than the markers) along with the estimated critical temperature in the thermodynamic limit. The displayed MCMC critical temperature was extracted from a fit using the same temperature grid and sizes for the sake of fair comparison.
    Lower panels: deviation between the flow and MCMC estimates. The gray ribbon shows one standard error of the MCMC estimate, the thick error bars on the markers are single standard error bars of the flow's estimate, while the thin bar is the combined error bar following the standard quadrature rule.}
    \label{fig:flow_binder}
\end{figure*}

On the second order side of the phase boundary, we perform a scan across $T$ at $\delta=1$ with $L=16,24,32,40$ at $\alpha=8, 16$
\footnote{Because our goal is a rigorous assessment of the effectiveness of flow models in comparison with existing Monte Carlo methodologies for
the $\phi^6$ model, we not only use the more sophisticated MCMC configuration generators described in the text, we also use larger lattice sizes than in the $\phi^4$ case
as well as the more precise fitting form for the Binder crossings of Eq.~\ref{eq:TcLaLb}.}.
Estimating the Binder cumulant of the magnetization $m = L^{-2}\sum_i \phi_i$, using flows, we find the critical temperature by inspecting crossings of adjacent sizes.
The expected finite size crossing form for $T_c$ is~\cite{Ballesteros_1996}
\begin{equation}
    T_c(L_a, L_b) = T_c^\infty + A \frac{L_b^{-\omega} - L_a^{-\omega}}{L_b^{1/\nu} - L_a^{1/\nu}}
    \label{eq:TcLaLb}
\end{equation}
where $\nu$ is the correlation length critical exponent (2D Ising universality in our case), and $\omega$ is a correction-to-scaling exponent.
We perform our fits only on the sizes $L=24,32,40$ as the scaling form above does not describe $L=16$ at the precision of the data produced by the flows.
Due to the limited sizes we simulated, we cannot fit $\omega$ directly. 
So, although not completely theoretically justified, we set $\omega$ to one of the leading scaling correction exponents of the 2D Ising model, $7/4$ or $2$~\cite{Caselle_2002} for our fits, and find that both give fits of similar quality (in $\chi^2$) given how close together the two exponents are.
The limited range of $L$ values used for our fit is not large enough to resolve the differences between the two correction terms.
Hence, for our estimates of $T_c^\infty$ we report the average of the two.
For a fair comparison, we also extract an estimate of $T_c^\infty$ from MCMC data in the same way and using the same sizes; the Binder cumulants and resulting critical temperature estimates are shown together with the flow model estimates in \cref{fig:flow_binder}.
Although the results are not as striking as in the first-order case, we again see remarkably good agreement between the two methods.

\begin{figure*}
    \centering
    \includegraphics[width=\linewidth]{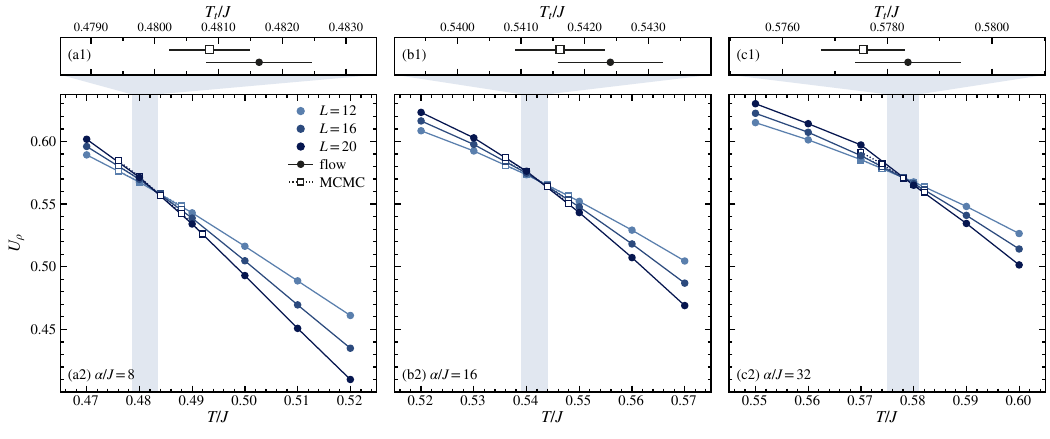}
    \caption{
    Lower panels:  Crossings of the Binder cumulant $U_\rho$ of the density restricted to the zero-skew curve for both flows and MCMC.
    Upper panels:  Temperature locations $T_{t}$ of the tricritical points.
    }
    \label{fig:flow_tricrit}
\end{figure*}

For the tricritical point, due to limitations on compute time, we focus on analyzing flow data for $\alpha=8,16,32$ and only $L=12,16,20$.
As explained in \cref{app:flow}, we use a mixture of three flow models, each restricted to specific density ranges, to simulate the tricritical point.
Unlike in the first-order case, where we are able to make use of both mixture components individually to detect the critical point, in the case of the tricritical point we found that using the same method as we had done in the MCMC case gave better estimates.
Just like in the MCMC case, we find the curve of zero skewness of the density histogram via reweighting in $\delta$. 
Along this curve we compute the Binder cumulant of the density, and perform fits to the scaling form \cref{eq:density_binder_scaling}.
We find that performing the fits while sharing $U_\rho^\ast$ and $r$ across the three different $\alpha$ values gives the cleanest fits; $U_\rho^\ast$ due to universality, and $r$ due to the fact that, in principle, it should be a ratio of universal scaling functions.
We show the results in \cref{fig:flow_tricrit} with comparison to MCMC references, plotting the Binder cumulants for the three sizes $L=12,16,20$ at the three $\alpha$ values.

\section*{Conclusions}

This work has three main ingredients.
First, we have introduced a particularly transparent soft-spin realization of the Blume–Capel model,
\cref{eq:BC},
in which the tuning parameter $\alpha$ directly controls how sharply the field is confined near $0,\pm1$, 
thereby making the connection between a continuous $\phi^6$ theory and the discrete Blume–Capel model explicit. 
In doing so we have also highlighted a mapping between the soft and discrete spin vacancy potentials 
which accounts for the detailed structure of the well curvatures.
Second, we have computed the finite-$\alpha$ phase diagrams, including the movement of the tricritical point 
as the model evolves from a soft continuous field toward the spin-1 limit. 
Third, we test the same normalizing-flow architecture as 
the character of the transition changes:
$ 
\text{Ising critical} \;\longrightarrow\;
\text{tricritical} \;\longrightarrow\; 
\text{first order}.
$
While flows have been tested at first-order transitions~\cite{Bai2024FlowFirstOrder}, 
we have extended this by addressing how flows' performance evolves as one crosses a tricritical point. 
In contrast,
the existing literature is more fragmented:  the
$\phi^4$ studies emphasize continuous criticality and critical slowing down, 
while the $\phi^6$ analyses focus on specialized first-order/nucleation sampling.

Overall, this work draws a complementary and reinforcing
statistical-mechanics picture about a transparent continuous interpolation to Blume–Capel physics, 
with an algorithmic component about how flow methods behave when the nature of the 
sampling bottleneck changes at tricriticality.

\section*{Acknowledgments}

E.M., E.K., and R.T.S.~were supported by the U.S. Department of Energy, Office of Science, Office
of Basic Energy Sciences, under Award Number DE-SC0022311 which constituted the primary
source of funding, in particular for the flow model studies. 
R.L.~contributed to the MCMC work and was supported by the U.S.~Department of Energy, Office of Science, Office of Basic Energy Sciences, under Award Number DE‐SC0014671.
E.M., E.K., and R.T.S.~acknowledge support in part by grant NSF PHY-2309135 to the Kavli Institute for Theoretical Physics (KITP), where part of their work was performed.
Computer simulations were made possible thanks San Jos{\'e} State University’s GPU cluster acquired through the NSF Grant No. OAC-2430291 as well as the Spartan high-performance computing facility at San Jos{\'e} State University acquired through the NSF Grant No.~OAC-1626645. 

\newpage

\bibliography{phi6}

\newpage

\appendix
\crefalias{section}{appendix}
\crefalias{subsection}{subappendix}
\crefalias{subsubsection}{subsubappendix}

\renewcommand{\thesection}{\Alph{section}}
\renewcommand{\thesubsection}{\arabic{subsection}}

\counterwithin*{figure}{section}
\counterwithin*{table}{section}
\counterwithin*{equation}{section}

\renewcommand{\thefigure}{\Alph{section}\arabic{figure}}
\renewcommand{\theHfigure}{\thefigure}
\renewcommand{\thetable}{\Alph{section}\arabic{table}}
\renewcommand{\theHtable}{\thetable}
\renewcommand{\theequation}{\Alph{section}\arabic{equation}}
\renewcommand{\theHequation}{\theequation}

\vskip0.50in

\section{Mean Field Theory}\label{app:mft}

\subsection{\bookmarkphi{4} model}\label{app:phi4_mft}

The one-site mean-field Hamiltonian corresponding to the continuous spin
version of the Ising model in \cref{eq:Ising} is
\begin{equation}
H_{\rm MF}(\phi)
=
\lambda\left(\phi^2-1\right)^2-zJm\phi.
\end{equation}
where
$m=\langle \phi\rangle$
and $z$ is the coordination number. For a square lattice, $z=4$.

The self-consistency equation is therefore
\begin{equation}
m=
\frac{
\displaystyle
\int_{-\infty}^{\infty}d\phi \,\,
\phi\,
\exp\left\{
-\beta\left[
\lambda\left(\phi^2-1\right)^2-zJm\phi
\right]
\right\}
}{
\displaystyle
\int_{-\infty}^{\infty}d\phi\,
\exp\left\{
-\beta\left[
\lambda\left(\phi^2-1\right)^2-zJm\phi
\right]
\right\}
}.
\end{equation}

Expanding the self-consistency equation to first order in \(m\), and using the fact that the \(m=0\) distribution is symmetric so that
$\langle \phi\rangle_0=0,$
one obtains,
\begin{equation}
\langle \phi\rangle_h
=
\beta h \langle \phi^2\rangle_0+O(h^3),
\end{equation}
where
This leads to,
$m= \beta zJm\,\langle\phi^2\rangle_0$,
and a nonzero infinitesimal value of $m$ becomes possible when
\begin{equation}
T_c=
zJ\langle\phi^2\rangle_{T_c}.
\label{eq:Tcmft}
\end{equation}
Eq.~\ref{eq:Tcmft} has a crisp interpretation:
While the unnormalized transition temperature $T_c$ depends on $\lambda$
through $\langle\phi^2\rangle_{T_c}$,
the transition temperature, when normalized by
$\langle\phi^2\rangle_{T_c}$, is equal to $zJ$ for all
$\lambda$, identical to the usual discrete spin Ising result.   

This result is more physically manifest in terms of the zero-field single-site
susceptibility. In the disordered phase,
\begin{equation}
\chi_0
=
\left.
\frac{\partial\langle\phi\rangle}{\partial h}
\right|_{h=0}
=
\beta\langle\phi^2\rangle.
\end{equation}
Then the mean-field instability criterion is
\begin{equation}
zJ\chi_0=1,
\end{equation}
which immediately reproduces \cref{eq:Tcmft}.

\begin{figure}[t!]
    \centering
    \includegraphics[width=\columnwidth]{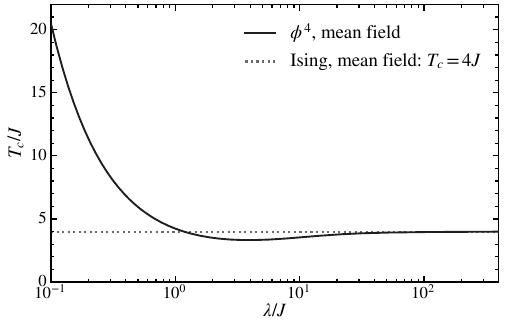}
    \caption{$T_c(\lambda)$ for the $\phi^4$ model within mean field theory.  
    Although  $T_{c}/\langle\phi^2\rangle_{T_c}$, is equal to $4J$ for all $\lambda$, the un-normalized $T_c$ shown here has a non-trivial
    $\lambda$ dependence.  Interestingly, it has the same shallow minimum found in the (exact) Monte Carlo
    results of \cref{fig:Tcphi4}.
    }
    \label{fig:Tcphi4mft}
\end{figure}

The MFT unnormalized $T_c(\lambda)$ is shown in \cref{fig:Tcphi4mft}
for the square lattice $z=4$.  
$T_c$ becomes very large as $\lambda \to 0^+$, since the spins become very soft;
$T_c/J \to 4$ in the hard-spin limit $\lambda \to \infty$, recovering the Ising mean-field result. 
In fact, an analytic result for $T_{c}$ in the $\lambda \rightarrow 0$ can be obtained
\begin{equation}
\frac{T_{c}}{J} = 16 \, \left\lbrack \frac{\Gamma(3/4)}{\Gamma(1/4)} \right\rbrack^{2} \, \frac{J}{\lambda} + 2.17  \,\, \simeq \,\,
1.83 \frac{J}{\lambda} + 2.17 J 
\end{equation}
so that $T_{c}$ diverges as $\lambda^{-1}$.
The shallow minimum $T_c < 4J$ at intermediate $\lambda/J$
is seen also in the exact Monte Carlo results of \cref{fig:Tcphi4}.

\begin{figure}[t!]
    \centering
     \includegraphics[width=\columnwidth]{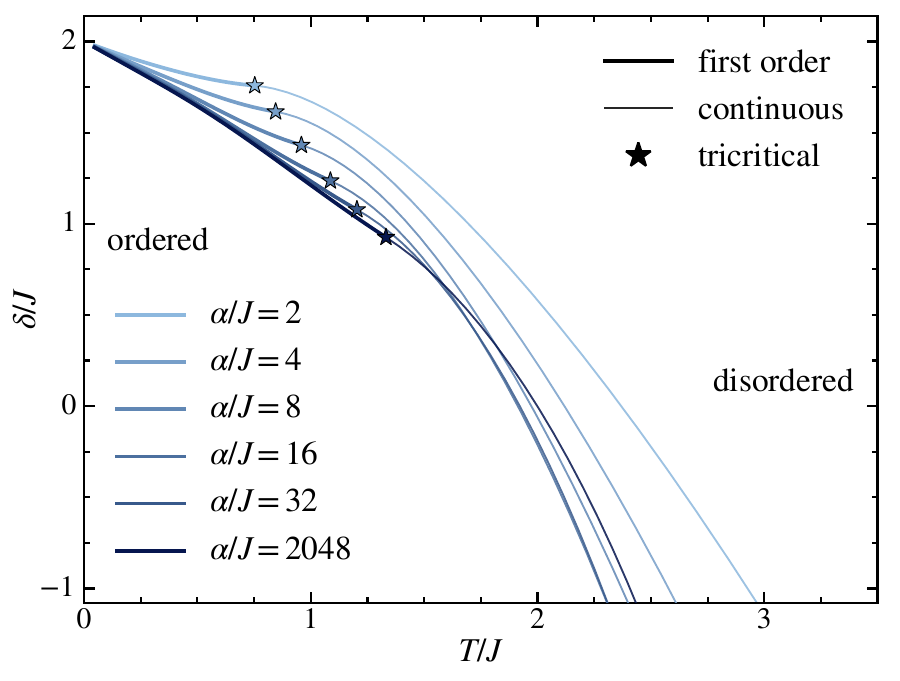}
    \caption{Phase diagram in the $T/J${--}$\delta/J$ plane of the continuous spin version of the Blume{--}Capel model for $\alpha/J=1,2,4,8$.
    The positions of the tricritical points are shown as solid circle markers.
    }
    \label{fig:Tcphi6mft}
\end{figure}

\subsection{\bookmarkphi{6} model}

In the soft-spin version of the Blume{--}Capel model, the effective one-site mean field Hamiltonian is,
\begin{equation}
H_{\rm MF}(\phi)
=
\alpha\,\phi^2(\phi^2-1)^2
+\delta\,\phi^2
-zJm\phi,
\end{equation}
where
$m=\langle\phi\rangle.$

As long as the transition is second order,
similar to  the $\phi^4$ (Ising) case,
for the square lattice the critical temperature
$T_c/\langle\phi^2\rangle_{T_c} = 4J$
and hence depends on $\alpha$ only through the length of the spin.

The tricritical point is obtained by solving simultaneously
\begin{equation}
T=
4J\langle\phi^2\rangle_0
\end{equation}
and
\begin{equation}
\langle\phi^4\rangle_0
=
3\langle\phi^2\rangle_0^2.
\end{equation}

\begin{table}[h!]
\centering
\begin{tabular}{ c|c|c }
 $\alpha/J$ & $T_t/J$ & $\delta_t/J$ \\
 \hline
 $1$ & $0.683$ & $1.857$ \\  
 $2$ & $0.753$ & $1.758$ \\  
 $4$ & $0.845$ & $1.615$ \\  
 $8$ & $0.959$ & $1.431$ \\  
 $16$ & $1.086$ & $1.236$ \\  
 $32$ & $1.204$ & $1.078$ \\  
 $2048$ & $1.332$ & $0.926$
\end{tabular}
\caption{Locations of the tricritical points of the soft-spin Blume-Capel predicted by Mean Field Theory.}
\label{tab:mft_tricritical}
\end{table}
    The positions of the tricritical points are tabulated in \cref{tab:mft_tricritical}.
For comparison, the discrete Blume{--}Capel model in mean-field theory on a square lattice has
     $(T_t/J,\Delta_t/J) =  (1.333,1.848)$, which, after applying the $T \ln 2$ shift discussed in \cref{ssec:bc_phi6}, gives $(T_t/J, \delta_t/J) = (1.333, 0.924)$ in agreement with the $\alpha=2048$ mean-field theory prediction.

\section{The anatomy of the flow model}\label{app:flow}

\subsection{Base Density}\label{sssec:base_density}
The flow maps samples from a tractable base density $p_b$ to field configurations~\cite{Albergo2019FlowBased,Albergo2021Introduction,papamakarios2021}. 
In order to reduce the amount of work the flow must do, we try to encode as much as we can about the target distribution into the base. 
In particular, we try to capture the modality of the target distribution in our base, as splitting a single mode into multiple modes can sometimes be challenging for invertible networks~\cite{hackett2025,Hagemann_2021,izmailov2019}.
As there are three wells in the field theory we study, we construct a three-component multi-dimensional Gaussian mixture for our base.
\begin{equation}
    p_b(z) = \pi_c \mathcal{N}(0, \Sigma_c) + \frac{1-\pi_c}{2}\left( \mathcal{N}(\mu \mathbf{1}, \Sigma_o) + \mathcal{N}(-\mu \mathbf{1}, \Sigma_o)\right)
\end{equation}
where $\pi_c \in \lbrack0, 1\rbrack$, $\mu \in \mathbb{R}_{\geq 0}$, and $\Sigma_c, \Sigma_o \in \mathbb{R}^{L^2 \times L^2}$ are in general trainable parameters, and $\mathbf{1} \in \mathbb{R}^{L^2}$ is the vector of all ones.
To make the density satisfy the symmetries of the lattice, the inverses of the covariance matrices are parametrized as diagonal matrices in Fourier space with elements:
$\lambda_{j}(\mathrm{k}) = \exp(\theta_{j,\text{orbit}(\mathrm{k})})$, where $j \in \{c, o\}$ and $\text{orbit}(\mathrm{k})$ gives the unique $D_4$ orbit of the lattice momentum $\mathrm{k}$, meaning we store $\theta$ for only a single representative of the momenta that are equivalent under the point group symmetry of the lattice.
The mean $\mu$ is initialized by minimizing the reverse KL divergence between the ordered component{---}that is, the mixture component which is not centered at zero{---}to the field theory action.
The spectra computed by this minimization tend to be too narrow and are therefore discarded and replaced with the free-field form $\lambda(\mathrm{k}) = m_0^2 + \sum_\nu 4\sin^2(k_\nu/2)$ where we use $m_0^2 = 1/4$; both the center and outer wells' spectra are initialized this way.
Unless otherwise specified, the mixture weight $\pi_c$ is fixed to $1/2$.

\subsection{Coupling Layers}
A simple way to construct such a mapping is through the use of \textit{couplings}~\cite{dinh2017density}, in which the input lattice is split into two pieces $A$ and $B$, and the variables are updated as:
\begin{equation}\label{eq:coupling}
g(z) = \begin{cases}
    z_A^\prime = z_A \\
    z_B^\prime = h(z_B; \vartheta(z_A))
\end{cases}
\end{equation}
where $\vartheta$ are parameters of the transformation $h$ and are produced by a neural network, and $h(z; \vartheta)$ is an element-wise function which is simple to invert in its primary input $z$ given $\vartheta$.
It is clear that the coupling can therefore be inverted easily.
Different couplings can be applied in sequence, each with a different partitioning of the lattice, allowing all lattice sites to be updated.
We use a fixed checkerboard decomposition of the square lattice into sub-lattices $A$ and $B$, and alternate the roles of the two sub-lattices in the coupling layers.
We refer to a pair of coupling layers, one which updates sub-lattice $A$ given $B$, $g^{(A)}$, and the other which updates $B$ given $A$, $g^{(B)}$, as a ``block'' which we denote $G(z) = g^{(B)}(g^{(A)}(z))$.
Thus, our flow model can be constructed as a sequence of coupling blocks $\{G_i\}$: $f(z) = G_{M}(G_{M-1}(\cdots G_1(z) \cdots ))$ and the change-of-variables formula in log-space gives:
\begin{equation}
    \log p_\theta(\phi) = \log p_b(z^{(0)}) - \sum_{i=1}^M \log\abs{\det(\eval{\pdv{G_i}{z}}_{z=z^{(i-1)}})}
\end{equation}
where $z^{(i)}$ is the output of the $i$th block, with $\phi = z^{(M)} = f(z^{(0)})$, and $z^{(0)}$ being the sample from the base density.
By increasing the number of coupling blocks, we can produce progressively richer transformations to the base density, allowing us to model increasingly complex distributions.

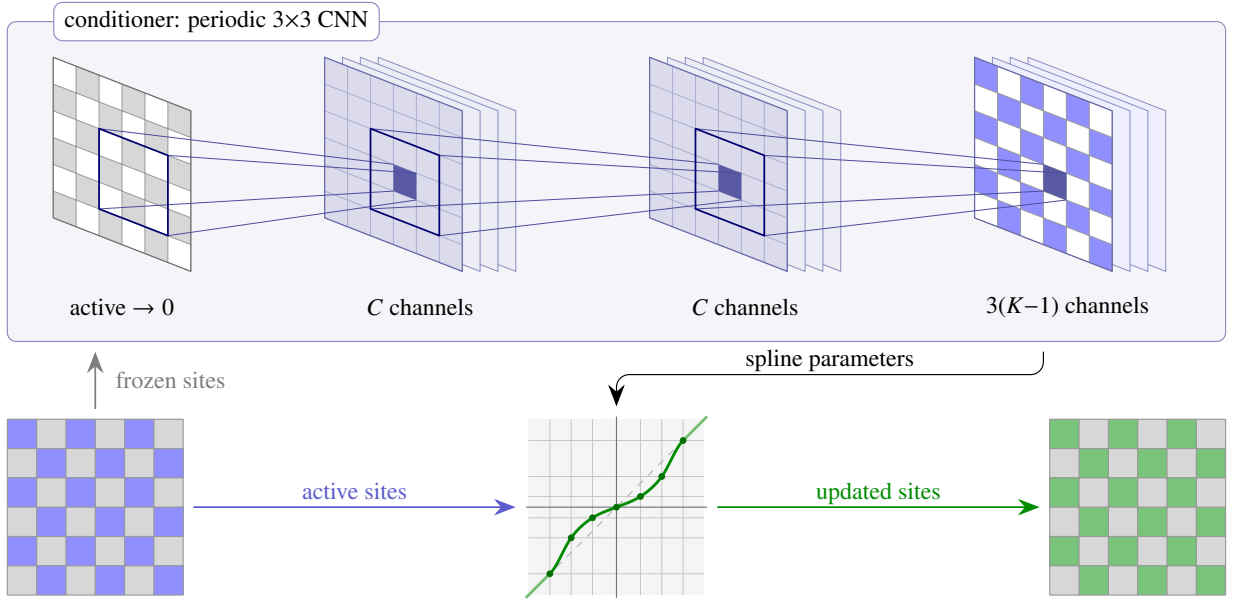
\begin{figure*}[btp!]
    \centering
    \providecommand{\couplingtextscale}{1}%
\ifstandalone\providecommand{\couplingwidth}{179mm}\else\providecommand{\couplingwidth}{\linewidth}\fi%
\pgfmathsetlengthmacro{\figunit}{\couplingwidth/96.0}%
\pgfmathsetlengthmacro{\figplotunit}{1.0455*\figunit}%
\pgfmathsetlengthmacro{\figarrow}{1.7*\figunit}%
\pgfmathsetlengthmacro{\figarrowwd}{1.3*\figunit}%
\pgfmathsetlengthmacro{\figcorner}{0.7*\figunit}%
\begin{tikzpicture}[
  x=\figunit,y=\figunit,
  font=\normalsize,
  >={Stealth[length=\figarrow,width=\figarrowwd]},
  lbl/.style={font=\normalsize,align=center,scale=\couplingtextscale},   %
  px1/.style={line width=0.24pt},
  px2/.style={line width=0.48pt},
  px3/.style={line width=0.72pt},
  plane/.style={shift={(#1,\ybase)},cm={\dvx,\dvy,0,\ch,(0,0)}},
]
\renewcommand{\baselinestretch}{1}   %
\colorlet{latFrozen}{cFrozen!35}
\colorlet{latActive}{cActive!80}
\colorlet{latUpdated}{cUpdated!52}
\colorlet{inkActive}{cActive!80!black}
\colorlet{inkUpdated}{cUpdated}
\def\cell{2.3}                 %
\def\rowmid{14.0}               %
\def\dvx{-1.8}\def\dvy{0.7}    %
\def\ch{2.1}                   %
\def\dc{1.4}                   %
\def\ybase{32.5}               %
\def\xin{14.4}\def\xha{35.7}\def\xhb{61.2}\def\xout{86.7}   %
\def\gap{0.8}                 %
\def\axend{6.8}\def\pfac{1.0455}               %
\def\xstk{81.3}\def\xs{47.8}\def\ye{24.4}                  %
\def\xy{81.8}                 %
\def\lblgap{0.65}              %

\newcommand{\onplane}[3]{({#1+(#2)*(\dvx)},{\ybase+(#2)*(\dvy)+(#3)*(\ch)})}

\newcommand{\board}[3]{%
  \begin{scope}[shift={(#1,\rowmid-3*\cell)}]
    \fill[#2] (0,0) rectangle (6*\cell,6*\cell);
    \foreach \ci in {0,...,5}{\foreach \cj in {0,...,5}{
      \pgfmathparse{int(mod(\ci+\cj,2))}
      \ifnum\pgfmathresult=1 \fill[#3] (\ci*\cell,\cj*\cell) rectangle ++(\cell,\cell);\fi}}
    \draw[black!40,px2,step=\cell] (0,0) grid (6*\cell,6*\cell);
  \end{scope}}

\newcommand{\backplanes}[2]{%
  \foreach \ck in {3,2,1}{
    \begin{scope}[plane={#1+\ck*\dc}]
      \filldraw[fill=#2,draw=cFlow!45,px1,line join=round] (0,0) rectangle (6,6);
    \end{scope}}}

\newcommand{\kernel}[2]{%
  \foreach \ia/\ja/\ib/\jb in {1/1/2/2, 4/1/3/2, 4/4/3/3, 1/4/2/3}
    \draw[cFlow!70,px1] \onplane{#1}{\ia}{\ja} -- \onplane{#2}{\ib}{\jb};
  \begin{scope}[plane={#1}]
    \draw[cFlow,px3,line join=round] (1,1) rectangle (4,4);
  \end{scope}}

\draw[rounded corners=\figcorner,fill=cFlow!4,draw=cFlow!45,px2] (0,27) rectangle (95.6,52.1);
\node[lbl,fill=white,draw=cFlow!45,px2,rounded corners=\figcorner,inner xsep=4.5pt,inner ysep=3.5pt,anchor=west] at (3.6,52.1)
  {conditioner: periodic $3{\times}3$ CNN};

\begin{scope}[plane={\xin}]
  \fill[white] (0,0) rectangle (6,6);
  \foreach \ci in {0,...,5}{\foreach \cj in {0,...,5}{
    \pgfmathparse{int(mod(\ci+\cj,2))}
    \ifnum\pgfmathresult=1 \fill[latFrozen] (\ci,\cj) rectangle ++(1,1);\fi}}
  \draw[black!40,px1,step=1] (0,0) grid (6,6);
  \draw[black!55,px2,line join=round] (0,0) rectangle (6,6);
\end{scope}

\foreach \xh in {\xha,\xhb}{
  \backplanes{\xh}{cFlow!7}
  \begin{scope}[plane={\xh}]
    \fill[cFlow!14] (0,0) rectangle (6,6);
    \draw[cFlow!32,px1,step=1] (0,0) grid (6,6);
    \draw[cFlow!60,px2,line join=round] (0,0) rectangle (6,6);
    \fill[cFlow!65] (2,2) rectangle (3,3);
  \end{scope}}

\backplanes{\xout}{cActive!12}
\begin{scope}[plane={\xout}]
  \fill[white] (0,0) rectangle (6,6);
  \foreach \ci in {0,...,5}{\foreach \cj in {0,...,5}{
    \pgfmathparse{int(mod(\ci+\cj,2))}
    \ifnum\pgfmathresult=0 \fill[latActive] (\ci,\cj) rectangle ++(1,1);\fi}}
  \draw[black!40,px1,step=1] (0,0) grid (6,6);
  \draw[cFlow!60,px2,line join=round] (0,0) rectangle (6,6);
  \fill[cFlow!65] (2,2) rectangle (3,3);
\end{scope}

\kernel{\xin}{\xha}
\kernel{\xha}{\xhb}
\kernel{\xhb}{\xout}

\node[lbl] at (9,29.7) {active $\to 0$};
\node[lbl] at (32.4,29.7) {$C$ channels};
\node[lbl] at (57.9,29.7) {$C$ channels};
\node[lbl] at (83.2,29.7) {$3(K{-}1)$ channels};

\board{0}{latFrozen}{latActive}

\draw[->,cFrozen!90!black,px3] (3*\cell,\rowmid+3*\cell+\gap) -- (3*\cell,27-\gap);
\node[lbl,cFrozen!90!black,anchor=west] at (3*\cell+1,{(\rowmid+3*\cell+\gap+26.2)/2}) {frozen sites};

\begin{scope}[shift={(\xs,\rowmid)},x=\figplotunit,y=\figplotunit]
  \fill[black!4] (-6.6,-6.6) rectangle (6.6,6.6);
  \foreach \kx in {-5,-3.4,-1.8,1.8,3.4,5} \draw[black!22,px1] (\kx,-6.6) -- (\kx,6.6);
  \foreach \ky in {-5,-2.3,-0.8,0.8,2.3,5} \draw[black!22,px1] (-6.6,\ky) -- (6.6,\ky);
  \draw[black!55,px2] (-\axend,0) -- (\axend,0);   %
  \draw[black!55,px2] (0,-\axend) -- (0,\axend);
  \draw[dashed,black!30,px2] (-\axend,-\axend) -- (\axend,\axend);
  \draw[very thick,inkUpdated!55] (-\axend,-\axend) -- (-5,-5);
  \draw[very thick,inkUpdated!55] (5,5) -- (\axend,\axend);
  \draw[very thick,inkUpdated,line join=round]
    plot coordinates {(-5.000,-5.000) (-4.900,-4.892) (-4.800,-4.768) (-4.700,-4.629) (-4.600,-4.474) (-4.500,-4.306) (-4.400,-4.127) (-4.300,-3.939) (-4.200,-3.744) (-4.100,-3.545) (-4.000,-3.346) (-3.900,-3.150) (-3.800,-2.960) (-3.700,-2.778) (-3.600,-2.607) (-3.500,-2.447) (-3.400,-2.300) (-3.300,-2.164) (-3.200,-2.037) (-3.100,-1.917) (-3.000,-1.804) (-2.900,-1.696) (-2.800,-1.595) (-2.700,-1.498) (-2.600,-1.407) (-2.500,-1.319) (-2.400,-1.236) (-2.300,-1.156) (-2.200,-1.079) (-2.100,-1.005) (-2.000,-0.934) (-1.900,-0.866) (-1.800,-0.800) (-1.700,-0.737) (-1.600,-0.679) (-1.500,-0.624) (-1.400,-0.573) (-1.300,-0.524) (-1.200,-0.477) (-1.100,-0.433) (-1.000,-0.390) (-0.900,-0.348) (-0.800,-0.308) (-0.700,-0.269) (-0.600,-0.230) (-0.500,-0.192) (-0.400,-0.154) (-0.300,-0.116) (-0.200,-0.078) (-0.100,-0.039) (0.000,0.000) (0.100,0.039) (0.200,0.078) (0.300,0.116) (0.400,0.154) (0.500,0.192) (0.600,0.230) (0.700,0.269) (0.800,0.308) (0.900,0.348) (1.000,0.390) (1.100,0.433) (1.200,0.477) (1.300,0.524) (1.400,0.573) (1.500,0.624) (1.600,0.679) (1.700,0.737) (1.800,0.800) (1.900,0.866) (2.000,0.934) (2.100,1.005) (2.200,1.079) (2.300,1.156) (2.400,1.236) (2.500,1.319) (2.600,1.407) (2.700,1.498) (2.800,1.595) (2.900,1.696) (3.000,1.804) (3.100,1.917) (3.200,2.037) (3.300,2.164) (3.400,2.300) (3.500,2.447) (3.600,2.607) (3.700,2.778) (3.800,2.960) (3.900,3.150) (4.000,3.346) (4.100,3.545) (4.200,3.744) (4.300,3.939) (4.400,4.127) (4.500,4.306) (4.600,4.474) (4.700,4.629) (4.800,4.768) (4.900,4.892) (5.000,5.000)};
  \foreach \kx/\ky in {-5/-5,-3.4/-2.3,-1.8/-0.8,0/0,1.8/0.8,3.4/2.3,5/5}
    \filldraw[inkUpdated!80!black] (\kx,\ky) circle (0.215);
\end{scope}

\draw[->,inkActive,px3] (6*\cell+\gap,\rowmid) -- node[lbl,anchor=base,yshift=\lblgap*\figunit,pos=0.5,inkActive] {active sites}
  (\xs-\axend*\pfac-\gap,\rowmid);
\draw[->,px2,rounded corners=1.0*\figunit] (\xstk,27-\gap) -- (\xstk,\ye) -- (\xs,\ye) -- (\xs,\rowmid+\axend*\pfac+\gap);
\node[lbl,anchor=base,yshift=\lblgap*\figunit] at ({(\xs+\xstk)/2},\ye) {spline parameters};

\board{\xy}{latFrozen}{latUpdated}
\draw[->,inkUpdated,px3] (\xs+\axend*\pfac+\gap,\rowmid) -- node[lbl,anchor=base,yshift=\lblgap*\figunit,pos=0.5,inkUpdated] {updated sites}
  (\xy-\gap,\rowmid);

\end{tikzpicture}
    \caption{
        A single coupling layer, mapping the field configuration on the left to the one on the right. The lattice is split by a checkerboard mask into frozen (gray) and active (blue) sites. The conditioner network sees the lattice with all active sites zeroed out and outputs parameters of rational-quadratic splines, which then take the active sites as input and produce the updated field values for those sites (green).
    }
    \label{fig:coupling_layer}
\end{figure*}

In our models, we use spline couplings~\cite{durkan2019neuralsplineflows} which define the functions transformations $h(z; \vartheta)$ using rational quadratic (RQ) splines. The conditioner network is a convolutional neural network whose responsibility is to generate the parameters $\vartheta$ of the RQ splines (the positions of the knots and their slopes) given the states of the frozen sites.
A schematic of our coupling layer is shown in \cref{fig:coupling_layer}.

\subsection{Multiscale Architecture}
In order to more efficiently encode the long-range structure of the field theory, we use a coarse-to-fine mapping inspired by~\cite{yu2020waveletflowfasttraining, wavelet_crg}. We draw a configuration $z$ from our $L\times L$ base density (see \cref{sssec:base_density}) and decompose each $2\times 2$ block by an orthonormal Haar transform into a block-mean, $z_c^\prime$, and three detail coefficients, $r_1, r_2, r_3$. 
The $L/2 \times L/2$ block-means are then passed through a stack of coupling layers as described above giving $z_c^\prime$.
The detail coefficients are then rescaled and shifted conditionally on $z_c^\prime$, giving $\zeta_i = e^{s_i(z_c^\prime)}r_i + t_i(z_c^\prime)$, similar to the standard RealNVP style affine transform~\cite{dinh2017density}.
An attempt was made to use RQ splines for this transformation as well but preliminary tests showed no benefit over the simpler transform.
We then use an inverse Haar transform to merge $z_c^\prime$ and $\zeta_i$ back into an $L\times L$ lattice.
One can consider this process to be a generalization of the up-sampling procedures used in super-resolving samplers as in~\cite{efthymiou2019,bauer2025} which copy data from a coarse lattice and use Gaussian noise to produce the extra degrees of freedom before applying further transformations.
In contrast, we produce our initial configurations from an $L\times L$ base density so that the sites in the $2\times 2$ block already have some amount of correlation.
After the inverse Haar transform, we apply further coupling layers on the $L\times L$ lattice to refine the field configurations.

\subsection{Defining the loss function} \label{sssec:loss}
We can train a flow model $p_\theta$ (where $\theta$ denote the parameters of the flow) by minimizing the (reverse) Kullback-Leibler (KL) divergence with respect to the probability density of the target distribution:
\begin{align}
    \operatorname{KL}(p_\theta \Vert p) &= \int \prod_i d\phi_i p_\theta(\phi) (\log p_\theta(\phi) - \log {p(\phi)}) \\
    &= \int \prod_i d\phi_i p_\theta(\phi) (\log p_\theta(\phi) + \beta E(\phi) + \log Z)\\
    &= \beta\ev{E}_{\phi \sim p_\theta} - S + \log Z
\end{align}
where we defined the entropy $S = -\ev{\log p_\theta}_{\phi\sim p_\theta}$ of our flow model.
Next, we note that the partition function of the target distribution is independent of the parameters of the flow model, and hence we may discard it from our loss function, giving:
\begin{equation}
    \mathcal{L}[p_\theta] = \beta\ev{E}_{\phi \sim p_\theta} - S.
\end{equation}
We note that minimizing this loss function is equivalent to minimizing the free energy of the distribution defined by the flow model.

The advantage of using the reverse KL divergence is that we do not need to train on any actual field configurations from the target distribution. 
However, one major drawback is the tendency for the reverse KL to cause mode-collapse, in which the model fixates on learning only a subset of modes of the data without covering the others.
This is quite inconvenient when one wishes to model symmetric densities or field theories near a first order transition where two competing phases coexist. 
Reverse KL training therefore carries the risk of the model learning only one of the symmetry broken phases well, or only one of the competing phases at coexistence.

\subsection{Symmetries}
In addition to collapse into a single symmetry broken phase, we could also imagine our flow model breaking translation or point-group symmetry.
For example, it may favor placing a large cluster of positive fields in one corner of the lattice instead of anywhere else.
Thankfully, preventing such cases is quite straightforward by constructing our flows to obey as many of the symmetries of the target distribution as possible.
There are three types of symmetries we can impose when modeling the $\phi^6$ theory: $\phi \to -\phi$ symmetry (which we'll refer to as $\mathbb{Z}_2$ symmetry), the $D_4$ point group symmetry of the square lattice, and translation symmetry.

The case of $\mathbb{Z}_2$ symmetry can be handled by modifying how we construct our spline transformations.
We want the transformation in \cref{eq:coupling} to be odd which is obviously satisfied by the frozen branch.
Making the active branch odd requires that $h(-z_B; \vartheta(-z_A)) = -h(z_B; \vartheta(z_A))$.
The RQ spline has $K$ bins and $K+1$ knots in the interval $[-B, B]$ and connects to a straight line outside of that interval.
For each bin, it needs $3$ numbers defining the change in $x$ and $y$, between the knot to its left and the knot to its right, along with the slope of the curve at the knot to its left, this gives $3K$ parameters.
At the ends of the interval we fix the knots to the points $(-B, -B)$ and $(B, B)$ as well as fixing their slopes to unity, this leaves $3(K-1)$ numbers for the conditioner network to produce, which then get mapped to the $3K$ spline parameters.
We will denote the widths, heights, and slope parameters of a spline as $(\Delta x_1,\ldots,\Delta x_K), (\Delta y_1, \ldots, \Delta y_K), (m_0, \ldots, m_K)$, with $m_0 = m_K = 1$.
Defining the output of the conditioner network as $\tilde{\vartheta}(z)$, the transformation is then constructed by first symmetrizing the spline parameters as $\vartheta(z) = (\tilde{\vartheta}(z) + R(\tilde{\vartheta}(-z))/2$, where $R$ is the transformation that reverses the spline parameters as $(\Delta x_i, \Delta y_i) \to (\Delta x_{K+1-i}, \Delta y_{K+1-i})$ and $\Delta m_i \to \Delta m_{K-i}$.
We then have $h(-z_B; \vartheta(-z_A)) = h(-z_B; R(\vartheta(z_A)) = -h(z_B; \vartheta(z_A))$, as desired.
We note that this is similar to the construction given in~\cite{DelDebbio2021TrivializingMaps}, however our formulation does not rely on a branching condition like theirs does, and is therefore a proper bijection and diffeomorphism.

The point group symmetry of the square lattice can be imposed much more easily.
Although one could manually symmetry average the $3\times3$ convolution kernels before applying them, we instead parametrize the kernels by storing one number per orbit instead of the full kernel, this means we only store $3$ numbers per kernel per channel.
By then implementing the convolution as simple sums and a matrix multiplication, we also manage to approximately halve the runtime of each training iteration relative to manual symmetry averaging of the kernels.
In order to match the full parameterization's training dynamics, we rescale the stored parameters by the square-root of their orbit size.
Symmetric kernels only make the CNN equivariant with respect to $D_4$ transformations centered on sites, not the lattice center due to the checkerboard decomposition breaking the equivalence of the two sub-lattices. 
This can be fixed by restoring translational symmetry.

One might expect that translation symmetry would be satisfied automatically once we require that the CNNs perform their convolutions with periodic boundaries, however as mentioned above, the in-equivalence of the two sub-lattices prevents this.
For coupling-only models, translation invariance is only broken on the level of a shift by one lattice site, and can easily be restored by defining our model density by the mixture:
\begin{equation}
    p_{\theta, \text{sym}}(\phi) = \frac{1}{2}\left( p_\theta(\phi) + p_\theta(t(\phi)) \right)
\end{equation}
where $t$ is the lattice translation operator which shifts the input by one lattice site.
We only need to include a single shift along one dimension in our symmetry average as the model is already invariant with respect to the point group.

For the multiscale flow model we use, the blocking procedure reduces the symmetry invariance of the model further.
Denote the translation group of the lattice by $T \cong \mathbb{Z}_L \times \mathbb{Z}_L$.
The blocked lattice's translation group on the other hand is isomorphic to $\mathbb{Z}_{L/2}\times\mathbb{Z}_{L/2}$, which is a subgroup of index four in $T$.
The checkerboard split of the coarse lattice brings another reduction by a factor of two, as only those translations which preserve the checkerboard colouring are preserved.
The (normal) subgroup of the surviving translations is thus: $T^\prime = \{(2a, 2b) \in T \;\vert\; a+b \equiv 0 \pmod 2\} \lhd T$, which has primitive vectors $(2,2)$ and $(2, -2)$.
The point group symmetry degrades as well to the normal subgroup which preserves checkerboard colouring $D_2 \lhd D_4$, giving another factor of two.
As a result, the model is invariant under $T^\prime \rtimes D_2$ in order to fully symmetrize our model we take:
\begin{equation}\label{eq:flow_sym}
    p_{\theta,\text{sym}}(\phi) = \frac{1}{\abs{D_4/D_2}}\sum_{r \in D_4/D_2} \frac{1}{\abs{T/T^\prime}} \sum_{t \in T/T^\prime} p_\theta(r(t(\phi)))
\end{equation}
which is a sum over 16 terms.
In our experiments, we only use the symmetrized density during evaluation and when computing reweighting factors, not during training as we found that most of the quality gain can be achieved by symmetrizing at evaluation time, and the remaining gap was not worth the multiplicative increase in training time.

\subsection{Mixtures of Walled Flows}\label{ssec:walled_flows}

Near the first order transition, the distribution becomes multimodal not just in the magnetization (which we solved by imposing $\mathbb{Z}_2$ symmetry), but in the density $\rho = \ev*{\phi^2}$ as well, due to coexistence between ordered (high density) and disordered (low density) phases.

As mentioned in \cref{sssec:loss}, the reverse KL loss is mode-seeking, meaning it will tend to focus on only a single phase in the coexistence regime.
In order to mitigate this undesirable behaviour, we instead train two flows in the first order region, each restricted to one of the phases.
Inspired by the tilt we used in our umbrella sampling simulations, we add a one-sided harmonic potential (which we refer to as a ``wall'') to the free-energy loss
\begin{align}
    \mathcal{L}[p_\theta \,\vert\, \rho \gtrsim \tfrac{1}{2}] &= \beta E(\phi) - S + k_o \lbrack \max(0, \tfrac{1}{2} - \rho)\rbrack^2 \\
    \mathcal{L}[p_\theta \,\vert\, \rho \lesssim \tfrac{1}{2}] &= \beta E(\phi) - S + k_d \lbrack \max(0, \rho - \tfrac{1}{2})\rbrack^2
\end{align}
where the $k$ values are the stiffness parameters of the potentials.
We set the stiffness for the disordered side to $k_d = 50$, while the ordered side's stiffness is scaled with the volume of the system as $k_o = 9 N$; these are hyper-parameters which we found worked well, we did not spend too much effort on tuning them.
Additionally, the disordered flow's base is set with its mixture weight frozen at $\pi_c = 1$ while the ordered base has $\pi_c = 0$.

Near the tricritical point, the two density peaks begin to merge, thus, we additionally introduce a third flow to model the intermediate density range:
\begin{equation}
\begin{aligned}
    \mathcal{L}[p_\theta \,\vert\, \rho_l \lesssim \rho \lesssim \rho_r] = \beta E(\phi) - S +& k_s(\lbrack \max(0, \rho_l - \rho)\rbrack^2 \\ &+ \lbrack \max(0, \rho - \rho_r)\rbrack^2)
\end{aligned}
\end{equation}
where we set $\rho_l = 0.25$ and $\rho_r = 0.65$ as these performed suitably well in our tests, again we did not spend too much effort tuning these hyper-parameters thoroughly.

Each flow in a pair or triple is trained independently and they are only brought together at evaluation time by forming the mixture:
\begin{equation}
    p_{\theta,\text{mix}}(\phi) = \sum_{m} w_m p_{\theta,m}(\phi)
\end{equation}
where the mixture weights are taken to be $w_o = w_d = \tfrac{1}{2}$ in for the first-order pairs and $w_o = w_d = 0.45, w_s = 0.1$.
Since the averaging is linear, this commutes with the symmetry averaging procedure of \cref{eq:flow_sym}, meaning it can be performed before or after the mixture average.

We note that our procedure is similar to the partitioning flow idea of~\cite{Bai2024FlowFirstOrder}, although we use a different functional form for our free-energy walls (one-sided harmonic potentials or squared ReLUs) while they use an exponential wall and additionally perform a more complex stitching procedure to combine their overlapping flow models in which they stitch the histograms produced by their flows, which is a procedure that can only be done as a post processing step, meaning one cannot  
We leave investigation of a more sophisticated stitching method for future work.

\subsection{Optimization}

Optimization of the reverse KL divergence was carried out using natural gradient descent~\cite{amari1998,martens2014} with a trust-region~\cite{schulman2015}.
We use the path gradient estimator of the reverse KL divergence~\cite{roeder2017,vaitl2022a,vaitl2022b} to reduce variance.
The per-sample scores $O_i = \nabla_\theta \ln p_\theta(\phi_i)$ are used to define the information metric $S = \sum_{i=1}^{N_s} O_i O_i^\top$, and the proposed change in model parameters is then computed by solving the system $(S + r\bar{d}\mathbb{I})\delta\theta = \nabla_\theta \mathcal{L}$ for $\delta\theta$.
Here we've defined $r$ which is the damping parameter in units of $\bar{d} = \Tr[S]/N_P$, the mean of the diagonal of $S$.
We do not need to actually form the metric explicitly as it is of size $N_P \times N_P$ (where $N_P$ is the number of parameters being optimized) and instead use the Woodbury identity to solve the usually much smaller $N_s \times N_s$ system.
The update is $\theta\to\theta-\gamma\,\delta\theta$ and we choose quantity $\gamma$ such that the gradient update does not produce a large change in the KL.
Specifically, the gradient update changes the KL{---}approximated to second order{---}by $\tfrac{1}{2} \gamma^2 (\delta\theta)^\top(S + r\bar d \mathbb{I})\delta\theta$.
Demanding that this change should not exceed a fixed radius $\kappa$, we take the largest $\gamma \leq 1$, that is, $\gamma=\min\{1,\sqrt{2\kappa/(\delta\theta^{\top}g)}\}$.
We allow $\kappa$ to scale with the system size, setting it to: $\kappa = 3\times10^{-4} L^2$.

\subsection{Estimating Physical Quantities}

Given enough representational capacity and training time, the flows should in principle be able to learn any distribution.
In practice, they will have some non-zero bias.
We can correct this bias through reweighting.
The expectation value with respect to the target distribution of a quantity $\mathcal{O}$ is given by:
\begin{equation}
    \ev{\mathcal{O}} = \int \dd{\phi} p(\phi)\mathcal{O}(\phi) = \frac{\int \dd{\phi} e^{-\beta E(\phi)}\mathcal{O}(\phi)}{\int \dd{\phi} e^{-\beta E(\phi)}}
\end{equation}
If we wish to use draws from $p_\theta$ instead of $p$, we can multiply and divide both the numerator and denominator by $p_\theta$, giving:
\begin{equation}
    \ev{\mathcal{O}} = \frac{\int \dd{\phi} p_\theta(\phi) \frac{e^{-\beta E(\phi)}}{p_\theta(\phi)}\mathcal{O}(\phi)}{\int \dd{\phi} p_\theta(\phi) \frac{e^{-\beta E(\phi)}}{p_\theta(\phi)}} 
    = \frac{\ev{w\mathcal{O}}_{p_\theta}}{\ev{w}_{p_\theta}}
\end{equation}
where we've defined $w(\phi) = e^{-\beta E(\phi)}/p_\theta(\phi)$.
The reweighted estimator is exact provided $\supp(p_\theta) \supseteq \supp(p)$.

\end{document}